\documentclass[aip,preprint,jcp]{revtex4-2}

\usepackage{bm} 
\usepackage{amssymb}
\usepackage{amsmath}
\usepackage{braket}

\usepackage{tikz}
\usepackage{subcaption}

\begin{document}

\title{Singlet Geminal Wavefunctions}
\author{Paul A. Johnson}
 \email{paul.johnson@chm.ulaval.ca}
\author{Jean-David Moisset}
\author{Marianne Gratton}
\author{\'{E}mile Baril}
\author{Marc-Antoine Plourde}
\author{Mathis Lefebvre}
\author{Marianne Kerleaux}
 \affiliation{D\'{e}partement de chimie, Universit\'{e} Laval, Qu\'{e}bec, Qu\'{e}bec, G1V 0A6, Canada}
\author{Paul W. Ayers}
 \affiliation{Department of Chemistry and Chemical Biology, McMaster University, Hamilton, Ontario, L8S 4M1, Canada}

\author{Patrick Cassam-Chena\"{i}}
 \affiliation{Universit\'{e} C\^{o}te d'Azur, CNRS, LJAD, UMR 7351, 06100 Nice, France}

\author{Stijn De Baerdemacker}
 \affiliation{Department of Chemistry, University of New Brunswick, Fredericton, New Brunswick, E3B 5A3, Canada}

\author{Dimitri Van Neck}
 \affiliation{Center for Molecular Modeling, Ghent University, Technologiepark 46, 9052 Zwijnaarde, Belgium}

\date{\today}

\begin{abstract}
Wavefunction forms based on products of electron pairs are usually constructed as closed-shell singlets, which is insufficient when the molecular state has a nonzero spin or when the chemistry is determined by $d$- or $f-$electrons.  
A set of two-electron forms are considered as explicit couplings of second-quantized operators to open-shell singlets. Geminal wavefunctions are constructed and their structure is elaborated.  Numerical results for small model systems clearly demonstrate improvement over closed-shell singlet pairs.
\end{abstract}

\maketitle

\section{Introduction}
Quantum chemistry is dominated by the idea of 1-electron states (orbitals). An optimal choice is made for the ground state Slater determinant, upon which a correction is added either empirically or based on contributions from low-lying excitations. Such an approach is correct if the reference Slater determinant is a good first approximation to the system under study. 
Many chemical systems require more than one Slater determinant for even a qualitatively correct description. Such systems are said to be strongly correlated. Examples include the breaking of chemical bonds (the effect is amplified for multiple bonds), reaction transition states, and transition metal systems. Whenever it is not possible to identify a spin-orbital as occupied or virtual, there is not a single dominant Slater determinant.

Wavefunction models based on antisymmetric products of 2-electron functions \emph{geminals} are a better starting point for strongly correlated systems.\cite{fock:1950,PG_27,mcweeny:1960,mcweeny:1961,PG_21} Up to now, most studies have been restricted to geminals that are closed-shell singlets, the most general case being the antisymmetrized product of interacting geminals (APIG).\cite{PG_36,PG_35,PG_29,silver:1970} Computations with APIG are \emph{not} feasible, though there are specific cases which are.\cite{moisset:2022a} First, one can make the geminals sparse: by restricting each spatial orbital to a specific geminal, one obtains the antisymmetrized product of strongly-orthogonal geminals (APSG).\cite{PG_17,PG_18,PG_28,PG_30,PG_32} Further, if each geminal contains at most two spatial orbitals, this becomes the generalized valence-bond / perfect pairing (GVB-PP)\cite{PG_20,PG_26,PG_31,PG_61,PG_62} approach. Next, if all the geminals are the same, one obtains the antisymmetrized geminal power (AGP).\cite{PG_38,PG_39,PG_75,PG_76,PG_77,PG_78,PG_79,PG_80,henderson:2019,khamoshi:2019,henderson:2020a,dutta:2020,harsha:2020,khamoshi:2020a,khamoshi:2021} By structuring the geminals, one obtains the eigenvectors of the reduced Bardeen-Cooper-Schrieffer (BCS) Hamiltonian,\cite{bardeen:1957a,bardeen:1957b} the so-called Richardson\cite{richardson:1963,richardson:1964,richardson:1965}-Gaudin\cite{gaudin:1976} (RG) states. RG states may be used as a basis for the wavefunction like Slater determinants for weakly correlated systems. The variational mean-field is feasible,\cite{johnson:2020,fecteau:2020a,fecteau:2020b,johnson:2021,fecteau:2022,faribault:2022,johnson:2023,johnson:2024a,johnson:2024b} and systematic improvement is possible with a configuration interaction (CI) of RG states. Epstein-Nesbet perturbation theory gives the same result at a \emph{much} reduced cost. Finally, by choosing one dominant orbital in each geminal, one obtains the antisymmetrized product of 1-reference orbital geminals (AP1roG)\cite{AP1roG,APr2G} or equivalently pair coupled cluster doubles (pCCD).\cite{stein:2014} This last approach is not feasible variationally, but is easily solved as a projected Schr\"{o}dinger equation (pSE).\cite{boguslawski:2014a,boguslawski:2014b,boguslawski:2014c,boguslawski:2015,boguslawski:2016a,boguslawski:2016b,boguslawski:2017,boguslawski:2019,boguslawski:2021}

For APIG, and all of its descendants, there are no unpaired electrons and are hence said to have zero \emph{seniority}. They are necessarily approximations to the doubly occupied configuration interaction (DOCI) which, as one might expect, is a CI of seniority-zero Slater determinants.\cite{weinhold:1967a,weinhold:1967b,veillard:1967,clementi:1967,cook:1975,carbo:1977} As orbital rotations will change the seniority of a given spatial orbital, DOCI and APIG must be orbital-optimized. 

The seniority-zero restriction will need to be lifted to treat systems breaking multiple bonds.\cite{bytautas:2011,alcoba:2013,alcoba:2014,henderson:2014,bytautas:2015,wahlen:2017,bytautas:2018,kossoski:2022,debaerdemacker:2024} The purpose of this contribution is to explore the more general electron-pair structures, in particular for open-shell singlets. A few approaches to this problem have been considered. There is the geminal mean-field configuration interaction (GMFCI) which may be iterated until convergence giving the geminal self-consistent field (GSCF) approach with various orthogonality constraints.\cite{cassam:2006,cassam:2007,cassam:2010} There are also the recently proposed 2D-block geminals.\cite{cassam:2023} Coupled-cluster doubles can be restricted to only singlet-excitations, giving a method called CCD0.\cite{bulik:2015} Finally, APsetG\cite{johnson:2017} is a geminal product wavefunction which is engineered to include open-shell possibilities despite the fact that it is algebraically a closed-shell geminal product. 

This manuscript is structured as follows. Section \ref{sec:lie} summarizes the Lie algebras obtained by coupling electrons to singlets in groups of 2 spatial orbitals (so(5)), and in $N$ levels (sp(N)). Section \ref{sec:gem} presents all the results necessary to employ a pSE approach for sp(N) geminals. Finally, section \ref{sec:results} presents numerical results for 4 electron systems. 

\section{Electron Pairs} \label{sec:lie}

\subsection{Closed-shell singlet: su(2)} 

Second quantized operators $a^{\dagger}_{i\sigma}$ and $a_{i\sigma}$ which respectively create and remove a $\sigma$-spin electron in spatial orbital $i$ have the structure
\begin{align}
	[a^{\dagger}_{i\sigma},a_{j\tau}]_+ = \delta_{ij}\delta_{\sigma\tau}.
\end{align}
They may be coupled together to create structured \emph{pairs} of electrons. The simplest such structure is closed-shell singlets, built from the objects
\begin{align} \label{eq:su2_ops}
	S^+_i = a^{\dagger}_{i\uparrow}a^{\dagger}_{i\downarrow}, \quad 
	S^-_i = a_{i\downarrow}a_{i\uparrow}, \quad
	S^z_i = \frac{1}{2} \left( a^{\dagger}_{i\uparrow}a_{i\uparrow} + a^{\dagger}_{i\downarrow}a_{i\downarrow} - 1 \right).
\end{align}
$S^+_i$ creates a pair of electrons in spatial orbital $i$, $S^-_i$ removes a pair of electrons from spatial orbital $i$, and $S^z_i$ counts the number of pairs in spatial orbital $i$: $+\frac{1}{2}$ (1 pair) or $-\frac{1}{2}$ (0 pairs). These objects have simply verified commutators
\begin{subequations}\label{eq:su2_directsum_structure}
	\begin{align}
		\left[ S^+_i , S^-_j     \right] &= 2 \delta_{ij} S^z_i \\
		\left[ S^z_i , S^{\pm}_j \right] &= \pm \delta_{ij} S^{\pm}_i.
	\end{align}
\end{subequations}
Each set of 3 operators $\{S^+_i,S^-_i, S^z_i \}$ is associated with a specific spatial orbital where the structure \eqref{eq:su2_directsum_structure} is the Lie algebra su(2). Operators associated to distinct spatial orbitals commute, and the complete structure is referred to as a collection of ``su(2) copies''.  

\subsection{Singlets in 2 orbitals: so(5)}
It is impossible to have an open-shell singlet in a single spatial orbital. In two spatial orbitals (labelled 1 and 2), we have two copies of su(2) defined in the same way as above. Taken together, these objects close the Lie algebra so(4) $\equiv$ su(2) $\oplus$ su(2), with structure constants given by eqns. \eqref{eq:su2_directsum_structure}. Defining an open-shell singlet requires \emph{four} operators 
\begin{subequations} \label{eq:sing_bispin}
\begin{align}
A_{\frac{1}{2}\frac{1}{2}} &= \frac{1}{2} \left( a^{\dagger}_{1\uparrow} a^{\dagger}_{2\downarrow} - a^{\dagger}_{1\downarrow}a^{\dagger}_{2\uparrow} \right) \\
A_{-\frac{1}{2}-\frac{1}{2}} &= -\frac{1}{2} \left( a_{2\downarrow} a_{1\uparrow} - a_{2\uparrow}a_{1\downarrow} \right) \\
A_{\frac{1}{2}-\frac{1}{2}}  &= -\frac{1}{2} \left( a^{\dagger}_{1\uparrow} a_{2\uparrow} + a^{\dagger}_{1\downarrow}a_{2\downarrow} \right) \\
A_{-\frac{1}{2}\frac{1}{2}}  &= -\frac{1}{2} \left( a^{\dagger}_{2\uparrow} a_{1\uparrow} + a^{\dagger}_{2\downarrow}a_{1\downarrow} \right)
\end{align}
\end{subequations}
to close a Lie algebra. Here, $A_{\frac{1}{2}\frac{1}{2}}$ creates an open-shell singlet across the two spatial orbitals while $A_{-\frac{1}{2}-\frac{1}{2}}$ removes an open-shell singlet. The other two objects are singlet-excitation operators.\cite{helgaker_book} The subscript labelling simplifies the structure constants, the non-zero ones being:
\begin{subequations}
\begin{align}
[S^z_1,A_{\sigma\tau}] &= \sigma A_{\sigma\tau}, \;\;\; [S^z_2,A_{\sigma\tau}] = \tau A_{\sigma\tau} \\
[S^{\pm}_1,A_{\sigma\tau}] &= \sqrt{\left(\frac{1}{2}\mp \sigma \right) \left( \frac{1}{2}\pm \sigma +1 \right)} A_{(\sigma\pm 1)\tau} \\
[S^{\pm}_2,A_{\sigma\tau}] &= \sqrt{\left(\frac{1}{2}\mp \tau \right) \left( \frac{1}{2}\pm \tau +1 \right)} A_{\sigma(\tau\pm 1)}
\end{align}
\end{subequations}
The commutators between these elements are given in Table \ref{ta:sing_bisp}. 

\begin{table}
\def\arraystretch{1.5}
\centering
\begin{tabular}{l | cccc}
$\left[ - , - \right]$ & $A_{-\frac{1}{2}-\frac{1}{2}}$ &$A_{\frac{1}{2}-\frac{1}{2}}$&$A_{-\frac{1}{2}\frac{1}{2}}$&$A_{\frac{1}{2}\frac{1}{2}}$\\ \hline
$A_{-\frac{1}{2}-\frac{1}{2}}$ & 0 & $\frac{1}{2}S^{-}_{2}$&$\frac{1}{2}S^{-}_{1}$&$\frac{1}{2}\left(S^{z}_{1} + S^{z}_{2} \right)$ \\
$A_{\frac{1}{2}-\frac{1}{2}}$ & $-\frac{1}{2}S^{-}_{2}$ &0& $\frac{1}{2}\left(S^{z}_{1}-S^{z}_{2} \right)$ & $-\frac{1}{2}S^{+}_{1}$ \\
$A_{-\frac{1}{2}\frac{1}{2}}$ & $-\frac{1}{2}S^{-}_{1}$ & $\frac{1}{2}\left( S^{z}_{2}-S^{z}_{1} \right)$ & 0 & $-\frac{1}{2}S^{+}_{2}$\\
$A_{\frac{1}{2}\frac{1}{2}}$ & $-\frac{1}{2}\left(S^{z}_{1} + S^{z}_{2} \right)$ & $\frac{1}{2}S^{+}_{1}$ & $\frac{1}{2}S^{+}_{2}$ &0\\
\end{tabular}
\caption{Structure constants of the open-shell singlet operators: elements in the left column appear \emph{first} in the commutator.}
\label{ta:sing_bisp}
\end{table}
These ten operators close the Lie algebra so(5), and it is instructive to classify the irreducible representations (irreps) according to the labels of the group chain
\begin{align*}
\underset{[l_1 l_2]}{SO(5)} \supset SO(4) \cong \underset{J_1}{SO(3)_1} \times \underset{J_2}{SO(3)_2} \supset \underset{M_1}{SO(2)_1} \times \underset{M_2}{SO(2)_2}.
\end{align*}
The chosen $SO(5)$ group labels denote the values of the Cartan elements on the highest weight in each irrep, sorted so that $l_1 > l_2$. Other choices are possible \cite{RN985}. The 16 possible states in two spatial orbitals are split into six irreps, see Table \ref{ta:so5sing}. 

\begin{table} 
\def\arraystretch{1.3}
\centering
\begin{tabular}{c|c|rr|r}
$[l_1,l_2]$ & $(J_1,J_2)$ & $M_1$ & $M_2$ \\
\hline
$[\frac{1}{2},\frac{1}{2}]$ & $(\frac{1}{2},\frac{1}{2})$ & $-\frac{1}{2}$ & $-\frac{1}{2}$ & $\ket{\theta}$ \\
& & $\frac{1}{2}$ & $-\frac{1}{2}$ & $S^+_1 \ket{\theta}$ \\
& & $-\frac{1}{2}$ & $\frac{1}{2}$ & $S^+_2 \ket{\theta}$ \\
& & $\frac{1}{2}$ & $\frac{1}{2}$ & $S^+_1 S^+_2 \ket{\theta}$ \\
& $(0,0)$ & $0$ & $0$ & $A_{\frac{1}{2}\frac{1}{2}} \ket{\theta}$ \\
\hline
$[\frac{1}{2},0]$ & $(\frac{1}{2},0)$ & $-\frac{1}{2}$ & $0$ & $a^{\dagger}_{2\uparrow}\ket{\theta}$ \\
 &  & $\frac{1}{2}$ & $0$ & $S^+_1 a^{\dagger}_{2\uparrow}\ket{\theta}$ \\
 & $(0,\frac{1}{2})$ & $0$ & $-\frac{1}{2}$ &  $a^{\dagger}_{1\uparrow}\ket{\theta}$ \\
 & & $0$ & $\frac{1}{2}$  & $S^+_2 a^{\dagger}_{1\uparrow}\ket{\theta}$ \\
\hline
$[\frac{1}{2},0]$ & $(\frac{1}{2},0)$ & $-\frac{1}{2}$ & $0$ & $a^{\dagger}_{2\downarrow}\ket{\theta}$ \\
 &  & $\frac{1}{2}$ & $0$ & $S^+_1 a^{\dagger}_{2\downarrow}\ket{\theta}$ \\
 & $(0,\frac{1}{2})$ & $0$ & $-\frac{1}{2}$ &  $a^{\dagger}_{1\downarrow}\ket{\theta}$ \\
 & & $0$ & $\frac{1}{2}$  & $S^+_2 a^{\dagger}_{1\downarrow}\ket{\theta}$ \\
\hline
$[0,0]$ & $(0,0)$ & $0$ & $0$ & $a^{\dagger}_{1\uparrow}a^{\dagger}_{2\uparrow}\ket{\theta}$ \\
\hline
$[0,0]$ & $(0,0)$ & $0$ & $0$ & $a^{\dagger}_{1\downarrow}a^{\dagger}_{2\downarrow}\ket{\theta}$ \\
\hline
$[0,0]$ & $(0,0)$ & $0$ & $0$ & $\frac{1}{2}\left(a^{\dagger}_{1\uparrow}a^{\dagger}_{2\downarrow} + a^{\dagger}_{1\downarrow}a^{\dagger}_{2\uparrow} \right)\ket{\theta}$
\end{tabular}
\caption{Irreps of so(5): $[\frac{1}{2},\frac{1}{2}]$ singlet, $[\frac{1}{2},0]$ doublet, $[0,0]$ triplet.}
\label{ta:so5sing}
\end{table}

\subsection{Singlets in $N$ orbitals: sp(N)}
Taking $N$ spatial orbitals together, we can couple second-quantized operators to produce all possible singlets. Each spatial orbital gets a copy of su(2), and each pair of spatial orbitals gets a four-element set like Eqs. \eqref{eq:sing_bispin}. Specifically, for each of the $\binom{N}{2}$ pairs of orbitals, with $i<j$, we will employ the elements
\begin{subequations} \label{eq:spn_operators}
\begin{align}
A^+_{ij} &= \frac{1}{2} 
	\left( 
		a^{\dagger}_{i\uparrow}a^{\dagger}_{j\downarrow} 
		- a^{\dagger}_{i\downarrow}a^{\dagger}_{j\uparrow}  
	\right) \\
A^-_{ij} &=  -\frac{1}{2} 
	\left(
		a_{j\downarrow}a_{i\uparrow} - a_{j\uparrow}a_{i\downarrow}
	\right) \\
A^0_{ij} &=  -\frac{1}{2} 
	\left(
		a^{\dagger}_{i\uparrow}a_{j\uparrow} + a^{\dagger}_{i\downarrow}a_{j\downarrow}
	\right) \\
A^0_{ji} &=  -\frac{1}{2} 
	\left(
		a^{\dagger}_{j\uparrow}a_{i\uparrow} + a^{\dagger}_{j\downarrow}a_{i\downarrow}
	\right).
\end{align}
\end{subequations}
Notice that
\begin{subequations}
\begin{align}
	\left( A^+_{ij}\right)^{\dagger} &= - A^-_{ij} \\
	\left( A^0_{ij}\right)^{\dagger} &= A^0_{ji}.   
\end{align}
\end{subequations}
With these definitions, the structure constants
\begin{subequations}
\begin{align}
[ A^+_{ij}, A^-_{i'j'}] &= \frac{1}{2}
	\left(
		\delta_{ii'} \delta_{jj'} + \delta_{ij'} \delta_{ji'} + \delta_{ii'}A^0_{jj'} + \delta_{jj'}A^0_{ii'} + \delta_{ij'}A^0_{ji'} + \delta_{ji'}A^0_{ij'}
	\right) \label{eq:spn+-} \\
[ A^0_{ij}, A^+_{i'j'}] &= -\frac{1}{2} \left( \delta_{ji'} A^+_{ij'} + \delta_{jj'}A^+_{ii'} \right) \label{eq:spn0+} \\
[ A^0_{ij}, A^-_{i'j'}] &=  \frac{1}{2} \left( \delta_{ii'} A^-_{jj'} + \delta_{ij'}A^-_{ji'} \right) \label{eq:spn0-} \\
[ A^0_{ij}, A^0_{i'j'}] &=  \frac{1}{2} \left( \delta_{ij'} A^0_{i'j} - \delta_{i'j}A^0_{ij'} \right)
\end{align}
\end{subequations}
are easily worked out. Two notes should be made regarding the structure constants. First, the term $\delta_{ij'}\delta_{ji'}$ in eq. \eqref{eq:spn+-} is an artifact of the chosen representation. Strictly speaking it should only appear from the commutator of elements like $[A^+_{21},A^-_{12}]$, but the object $A^+_{21}$ doesn't exist. Looking at the definition of $A^+_{12}$, we could define it similarly, and find that $A^+_{21} = A^+_{12}$. Second, the convention chosen for the signs of the objects \eqref{eq:spn_operators} has the result that the commutators \eqref{eq:spn0+} and \eqref{eq:spn0-} have signs opposite of the usual convention in su(2), where
$[S^z,S^{\pm}] = \pm S^{\pm}$.
Thus for $N$ spatial orbitals, the Lie algebra for pairs is sp(N). In particular, for $N=1$ the Lie algebras sp(1) $\cong$ su(2) are equivalent and the name su(2) is accepted. Similarly for $N=2$, sp(2) $\cong$ so(5), but these are only low-dimensional accidents: for $N>2$ the Lie algebras are no longer equivalent.
With the relevant Lie algebras understood and specified, we now move to many-body states.

\section{Geminal Products} \label{sec:gem}
The goal of this section is a pSE approach for a general singlet geminal wavefunction. The pSE will first be quickly summarised, along with the corresponding version for su(2). A pSE requires a collection of basis functions to project against, which in the case of su(2) is Slater determinants of doubly occupied orbitals. For sp(N), the natural basis is configuration state functions (CSFs), e.g. table \ref{ta:so5sing} for $N=2$.

\subsection{Projected Schr\"{o}dinger Equation}
Define a geminal as a two-electron state expressed in a given single-particle basis
\begin{align} \label{eq:gem_def}
G^{\dagger}_{\alpha} = \sum_{ij} g^{ij}_{\alpha} a^{\dagger}_i a^{\dagger}_j.
\end{align}
They are building blocks for states of $2M$ electrons in $N$ sites of the form
\begin{align} \label{eq:gem_prod}
\ket{\{g\}} = \prod^{M}_{\alpha =1} G^{\dagger}_{\alpha} \ket{\theta}.
\end{align} 
The Coulomb Hamiltonian
\begin{align} \label{eq:Coulomb}
	\hat{H}_C = \sum_{ij} h_{ij} \sum_{\sigma} a^{\dagger}_{i\sigma} a_{j\sigma} 
	+ \frac{1}{2} \sum_{ijkl} V_{ijkl} \sum_{\sigma \tau} a^{\dagger}_{i\sigma} a^{\dagger}_{k\tau} a_{l\tau} a_{j\sigma} + V_{NN}
\end{align}
is written in terms of the 1- and 2-electron integrals
\begin{align} \label{eq:1el_int}
	h_{ij} &= \int d\mathbf{r} \phi^*_i (\mathbf{r}) \left( - \frac{1}{2} \nabla^2 - \sum_I \frac{Z_I}{| \mathbf{r} - \mathbf{R}_I |} \right) \phi_j (\mathbf{r}) \\
	V_{ijkl} &= \int d\mathbf{r}_1 d\mathbf{r}_2 \frac{\phi^*_i(\mathbf{r}_1)  \phi_j(\mathbf{r}_1)  \phi^*_k(\mathbf{r}_2)  \phi_l(\mathbf{r}_2)  }{| \mathbf{r}_1 - \mathbf{r}_2|}. \label{eq:2el_int}
\end{align}
Chemists' notation is employed for the 2-electron integrals. With the wavefunction ansatz \eqref{eq:gem_prod}, the direct approach is to minimize the Rayleigh quotient
\begin{align}
	E[\{g\}] = \min_{\{g\}} \frac{\braket{\{g\} | \hat{H}_C | \{g\} }}{\braket{\{g\} | \{g\}}}
\end{align}
variationally. This requires the 1- and 2-body reduced density matrices (RDM) of $\ket{\{g\}}$, which are generally very expensive to compute. 

The same problem occurs with coupled-cluster theory, and the workaround is to employ the \emph{weak} formulation of the problem, the pSE.\cite{helgaker_book} Begin with the Schr\"{o}dinger equation
\begin{align}
	\hat{H}_C \ket{\{g\}} = E \ket{\{g\}}
\end{align}
and notice that the state \eqref{eq:gem_prod} can be written in a conveniently chosen basis $\{\ket{ \Phi_{\omega} } \}$
\begin{align}
	\ket{\{g\}} = \sum_{\omega} C_{\omega} \ket{\Phi_{\omega}}.
\end{align}
By projecting from the left with a basis vector $\bra{\Phi_{\mu}}$
\begin{align}
	\braket{\Phi_{\mu} | \hat{H}_C | \{g\} } = E \braket{\Phi_{\mu} | \{g\}}
\end{align}
and noting the action of the Coulomb Hamiltonian on the basis vectors
\begin{align}
	\hat{H}_C \ket{\Phi_{\mu}} = \sum_{\nu} \lambda_{\mu\nu} \ket{\Phi_{\nu}} + \ket{\Phi^{\perp}}.
\end{align}
With $\ket{\Phi^{\perp}}$ orthogonal to the vectors $\{\ket{ \Phi_{\omega} } \}$, one arrives at a pSE
\begin{align} \label{eq:pse}
	\sum_{\nu} \left( \lambda_{\mu\nu} - \delta_{\mu\nu}E \right) \sum_{\omega} C_{\omega} \braket{\Phi_{\nu} | \Phi_{\omega}} = 0.
\end{align}
Each chosen vector $\bra{\Phi_{\mu}}$ leads to one such eq. \eqref{eq:pse}, and enough must be chosen to solve for the variables: the coefficients $\{g\}$ and $E$. The information thus required is the matrix elements $\lambda_{\mu\nu}$, the expansion coefficients $C_{\omega}$, and the overlap matrix elements $\braket{\Phi_{\nu} | \Phi_{\omega}}$. The equations \eqref{eq:pse} are \emph{nonlinear} since the expansion coefficients $C_{\omega}$ are functions of the geminal coeficients $\{g\}$. The matrix elements $\lambda_{\mu\nu}$ are functions of the integrals \eqref{eq:1el_int} and \eqref{eq:2el_int}, while the overlap matrix elements are combinatorial factors. 

\subsection{Closed-Shell Singlets: su(2)}
We will first outline the structure for closed-shell singlets as it provides a starting point for our development. Restricting the geminal \eqref{eq:gem_def} to closed-shell singlets yields
\begin{align}
G^{\dagger}_{\alpha} [\text{su}(2)] = \sum^{N}_{i=1} g^i_{\alpha} S^+_i, \label{eq:su2_gem}
\end{align}
with $S^+_i$ defined as in \eqref{eq:su2_ops}. We now expand a many-body state 
\begin{align}
\ket{\textbf{su(2)},M} = \prod^{M}_{\alpha=1} \sum^N_i g^i_{\alpha} S^+_i \ket{\theta} = \sum_{\omega} C_{\omega} \ket{\Phi_{\omega}}, \label{eq:su2_state}
\end{align}
in terms of Slater determinants, $\ket{\Phi_{\omega}}$, as basis functions and evaluate the corresponding expansion coefficients $C_{\omega}$. Specifically, $\omega$ is an ordered set of $M$ integers such that $1 \leq \omega_1 \leq \omega_2 \leq \dots \leq \omega_M \leq N$, labelling which spatial orbitals are doubly occupied
\begin{align}
	\ket{\Phi_{\omega}} = S^+_{\omega_1} S^+_{\omega_2} \dots S^+_{\omega_M} \ket{\theta}.
\end{align}
As Slater determinants are orthonormal, the projection we care about is precisely the expansion coefficient
\begin{align}
\braket{\Phi_{\omega} | \textbf{su(2)},M} = \sum_{\omega '} C_{\omega '} \braket{\Phi_{\omega} | \Phi_{\omega '}} = C_{\omega}.
\end{align}
From here a projected Schr\"{o}dinger equation approach follows easily.

In each of the monomials of $S^+$'s in Eq. \eqref{eq:su2_state}, any $S^+_i$ can only occur \emph{once} since 
\begin{align}
S^+_i S^+_i = a^{\dagger}_{i \frac{1}{2}} a^{\dagger}_{i -\frac{1}{2}} a^{\dagger}_{i \frac{1}{2}} a^{\dagger}_{i -\frac{1}{2}} = 0.
\end{align}
Further, the ordering in each monomial does not matter since
\begin{align}
[ S^+_i , S^+_j ] =0.
\end{align}
The expansion coefficients are thus symmetric sums of a product of factors:
\begin{align}
C_{\omega} = \sum_{\pi \in \mathfrak{S}_M} \prod^M_{\alpha =1} g^{\omega_{\alpha}}_{\pi (\alpha)}.
\end{align}
The sum is performed over all $M!$ permutations $\pi$ of $M$ objects, the symmetric group $\mathfrak{S}_M$. The previous statement is just the Leibniz formula for the matrix permanent
\begin{align}
C_{\omega} = \left|
	\begin{array}{ccc}
	g^{\omega_1}_1 & & g^{\omega_M} \\
	& \ddots & \\
	g^{\omega_1}_M & & g^{\omega_M}_M 
	\end{array} \right|^+.
\end{align}
Because permanents are not invariant to linear transformations,\cite{minc:1978} they are in general intractable to compute; though there are some forms which reduce to tractable expressions.\cite{borchardt,muir:1897,carlitz:1960,slavnov:1989,belliard:2019}

Finally, the action of the Coulomb Hamiltonian on a closed-shell Slater determinant yields
\begin{align}
	\hat{H}_C \ket{\Phi_{\omega}} &= \sum_{i \in \omega} \left( 2 h_{ii} + \sum_{j \in \omega (\neq i)} (2V_{iijj} - V_{ijji}) + V_{NN} \right) \ket{\Phi_{\omega}} 
	+ \sum_a \sum_{i \in \omega} V_{aiai} S^+_a S^-_i \ket{ \Phi_{\omega} } + \ket{\Phi^{\perp}}.
\end{align}

\subsection{Singlets: sp(N)}
Geminals for singlets with an so(5) structure could be defined and studied, but the main purpose is to build states for singlets distributed over the entire space. The individual singlet creators are
\begin{align}
A^+_{ij} = \frac{1}{2} \left(
a^{\dagger}_{i\uparrow} a^{\dagger}_{j \downarrow} - a^{\dagger}_{i \downarrow} a^{\dagger}_{j \uparrow}
\right).
\end{align}
Since $A^+_{ii} = S^+_i$ and $A^+_{ij} = A^+_{ji}$, an appropriate geminal is
\begin{align}
G^{\dagger}_{\alpha} [ \text{sp}(N)] = \sum_{ij} g^{ij}_{\alpha} A^+_{ij} = \sum_i g^{ii}_{\alpha} S^+_i + 2 \sum_{i < j} g^{ij}_{\alpha} A^+_{ij}, \label{eq:spN_geminal}
\end{align}
where the geminal coefficients are symmetric $g^{ij}_{\alpha} = g^{ji}_{\alpha}$. (If these coefficients were treated as distinct objects, only the symmetric portion of their linear combination would contribute in final formulas.) A geminal product can be written in terms of CSFs with a symmetry label $\eta$ denoting the number of closed-shell pairs
\begin{align}
\ket{\textbf{sp(N)},M} = \prod^M_{\alpha = 1} G^{\dagger}_{\alpha} [ \text{sp}(N)] \ket{\theta} = \sum_{\eta} \sum_{\omega} C^{\eta}_{\omega} 
\ket{\Phi^{\eta}_{\omega}}. \label{eq:spN_state}
\end{align}
The labelling $\omega$ is now a set of $M$ pairs of index labels $(i,j)$. For example, the CSF $A^+_{i_1 , j_1} A^+_{i_2 ,j_2} A^+_{i_3 ,j_3} \ket{\theta}$ has $\omega = \{ (i_1,j_1),(i_2,j_2),(i_3,j_3)  \}$. There are $\eta$ repeated indices. Without any loss of generality, we can list the $\eta$ repeated indices first, such that
\begin{align}
\omega = \{
(i_1,i_1),\dots,(i_{\eta},i_{\eta}),(i_{\eta +1},j_{\eta +1}),\dots,(i_M,j_M)
\}. \label{eq:omega_example}
\end{align}
The projection of the state \eqref{eq:spN_state} onto CSFs is now more complicated because the CSFs do not form an orthogonal basis. In particular,
\begin{align}
\braket{ \Phi^{\eta '}_{\omega '} | \textbf{sp(N)},M } = \sum_{\eta} \sum_{\omega} C^{\eta}_{\omega} \braket{ \Phi^{\eta '}_{\omega '} | \Phi^{\eta}_{\omega} }
\end{align}
has contributions from both the expansion coefficient $C^{\eta}_{\omega}$, and the overlap $ \braket{ \Phi^{\eta '}_{\omega '} | \Phi^{\eta}_{\omega} }$. Both have internal structure. We will first deal with the expansion coefficients, then examine the CSF overlaps. For each, we will first look at a four-pair example (large enough to show the non-triviality) before presenting the general result.

When indices are shared across different open-shell pairs, they condense to closed-shell pairs, while no index can occur more than twice
\begin{align}
A^+_{ij} A^+_{ij} &= -\frac{1}{2} S^+_i S^+_j \label{eq:spn_struct1} \\
A^+_{ij} A^+_{ik} &= -\frac{1}{2} S^+_i A^+_{jk} \label{eq:spn_condensation} \\
A^+_{ij} S^+_i &= 0.
\end{align}
Creators acting on distinct sites commute
\begin{align}
[ A^+_{ij},A^+_{kl}] &= 0. \label{eq:spn_distinct}
\end{align}
Finally, there is a braiding of the elements $A^+_{ij}$ of the form
\begin{align}
A^+_{ij} A^+_{kl} + A^+_{ik} A^+_{jl} + A^+_{il}A^+_{jk} = 0 \label{eq:braiding}
\end{align}
which causes linear dependance in the CSF basis. This linear dependance may be dealt with either by orthogonalizing the CSF basis or by choosing a minimal set of non-orthogonal CSFs. 

As the Coulomb Hamiltonian has an exact representation in terms of the sp(N) generators
\begin{align}
	\hat{H}_C = -2 \sum_{ij} h_{ij} A^0_{ij} + \sum_{ijkl} V_{ijkl} (2 A^0_{ij} A^0_{kl} + \delta_{jk} A^0_{il}) + V_{NN},
\end{align}
its action on an open-shell singlet CSF is a linear combination of open-shell singlet CSFs with $\ket{\Phi^{\perp}} = 0$. In particular
\begin{align}
	\hat{H}_C \ket{ \Phi^{\eta}_{\omega} } &= \left( \sum_{i \in \omega} w_i + \sum_{i<j \in \omega} W_{ij} + V_{NN} \right) \ket{ \Phi^{\eta}_{\omega} }
	\nonumber \\
	&- 2\sum_{i (\neq j)} \sum_{j\in\omega} h_{ij} A^0_{ij} \ket{ \Phi^{\eta}_{\omega} } 
	+ \sum_{(i,k)\neq (j,l)} \sum_{j,l \in \omega} V_{ijkl} (2A^0_{ij}A^0_{kl} + \delta_{jk} A^0_{il} )\ket{ \Phi^{\eta}_{\omega} }
\end{align}
with the elements
\begin{align}
	w_i &= \begin{cases}
		2h_{ii} + V_{iiii}, \quad & \vert i\in\omega\vert = 2 \\
		h_{ii}, \quad & \vert i\in\omega\vert =1
	\end{cases}
\end{align}
\begin{align}
	W_{ij} &= \begin{cases}
		4V_{iijj} - 2 V_{ijji}, \quad & \vert i\in\omega\vert = \vert j\in\omega\vert = 2 \\
		2V_{iijj} - V_{ijji}, \quad & \vert i\in\omega\vert =1, \vert j\in\omega\vert=2 \;\text{or}\; \vert i\in\omega\vert=2,\vert j\in\omega\vert=1 \\
		V_{iijj} + V_{jjii}, \quad & \vert i\in\omega\vert = \vert j\in\omega\vert = 1, (i,j) \in \omega \\
		V_{iijj}, \quad & \vert i\in\omega\vert = \vert j\in\omega\vert = 1, (i,j) \notin \omega.
	\end{cases}
\end{align}
The singlet-excitation operators acting on $\ket{ \Phi^{\eta}_{\omega} }$ generate different CSFs. Because of the braiding \eqref{eq:braiding}, a minimal set must be chosen. A linearly independent, though non-orthogonal, choice is made with the use of Rumer diagrams.\cite{weyl:1932}

\subsubsection{Expansion Coefficients: Four Pair Example}

Consider the product
\begin{align}
\ket{\textbf{sp(N)},4} &= \prod^4_{\alpha=1} G^{\dagger}_{\alpha} [ \text{sp}(N)]\ket{\theta} \label{eq:4site_exp}
\end{align}
and look first at the expansion coefficient of a typical CSF with $\eta=0$, e.g.
\begin{align}
\ket{\Phi^{(0)}_{j_1 j_2 j_3 j_4}} &=
A^+_{i_1 j_1} A^+_{i_2 j_2} A^+_{i_3 j_3} A^+_{i_4 j_4} \ket{\theta}. \label{eq:4site_eta0}
\end{align}
To condense the notation for $\omega$, we will only note the second index in each pair. The CSF \eqref{eq:4site_eta0} is independent of the other CSFs in the expansion \eqref{eq:4site_exp}: any permutation of the indices $\{j_1,j_2,j_3,j_4\}$ yields a different CSF. Each $A^+_{i_k j_k}$ occurs in each of the four geminals $G^{\dagger}_{\alpha}$ with the contribution $2g^{i_k j_k}_{\alpha}$, and because the sp(N) singlet creators commute, eq. \eqref{eq:spn_distinct}, the coefficient is a symmetric sum weighted by $2^4$
\begin{align}
C^{(0)}_{j_1 j_2 j_3 j_4} &= 2^4
\left|
\begin{array}{cccc}
g^{i_1 j_1}_{1} & g^{i_2 j_2}_{1} & g^{i_3 j_3}_{1} & g^{i_4 j_4}_{1} \\
g^{i_1 j_1}_{2} & g^{i_2 j_2}_{2} & g^{i_3 j_3}_{2} & g^{i_4 j_4}_{2} \\
g^{i_1 j_1}_{3} & g^{i_2 j_2}_{3} & g^{i_3 j_3}_{3} & g^{i_4 j_4}_{3} \\
g^{i_1 j_1}_{4} & g^{i_2 j_2}_{4} & g^{i_3 j_3}_{4} & g^{i_4 j_4}_{4} 
\end{array}
\right|^+
= 2^4 \sum_{\pi \in \mathfrak{S}_4} \prod^4_{a=1} g^{i_a j_a}_{\pi (a)}. \label{eq:raw}
\end{align}
$\pi$ is one of the 4! permutations in the symmetric group $\mathfrak{S}_4$. 

Looking at a typical CSF with $\eta=1$ introduces a complication due to eq. \eqref{eq:spn_condensation}. The final coefficient $C$ of the CSF
\begin{align}
\ket{\Phi^{(1)}_{i_1 j_2 j_3 j_4}  } &=
S^+_{i_1} A^+_{i_2 j_2} A^+_{i_3 j_3} A^+_{i_4 j_4} \ket{\theta}
\end{align}
will receive contributions from the ``raw'' coefficients, $D$, from terms which are permutations of the index $i_1$, i.e.
\begin{align}
C^{(1)}_{i_1 j_2 j_3 j_4}
 S^+_{i_1} A^+_{i_2 j_2} A^+_{i_3 j_3} A^+_{i_4 j_4} \ket{\theta} &= 
D^{(1)}_{i_1 j_2 j_3 j_4}
 S^+_{i_1} A^+_{i_2 j_2} A^+_{i_3 j_3} A^+_{i_4 j_4} \ket{\theta}    \nonumber \\
&+ D^{(1)}_{j_2 i_1 j_3 j_4}
A^+_{i_1 j_2} A^+_{i_2 i_1} A^+_{i_3 j_3} A^+_{i_4 j_4} \ket{\theta} \nonumber \\
&+ D^{(1)}_{j_3 j_2 i_1 j_4} 
A^+_{i_1 j_3} A^+_{i_2 j_2} A^+_{i_3 i_1} A^+_{i_4 j_4} \ket{\theta} \nonumber \\
&+ D^{(1)}_{j_4 j_2 j_3 i_1} 
 A^+_{i_1 j_4} A^+_{i_2 j_2} A^+_{i_3 j_3} A^+_{i_4 i_1} \ket{\theta}.
\end{align}
The three additional terms arise from permutations of the index $i_1$, while permutations of the indices $j_k$ that do not involve $i_1$ generate \emph{different} CSFs. The first raw coefficient is 
\begin{align}
D^{(1)}_{i_1 j_2 j_3 j_4} &= 2^3
\left|
\begin{array}{cccc}
g^{i_1 i_1}_{1} & g^{i_2 j_2}_{1} & g^{i_3 j_3}_{1} & g^{i_4 j_4}_{1} \\
g^{i_1 i_1}_{2} & g^{i_2 j_2}_{2} & g^{i_3 j_3}_{2} & g^{i_4 j_4}_{2} \\
g^{i_1 i_1}_{3} & g^{i_2 j_2}_{3} & g^{i_3 j_3}_{3} & g^{i_4 j_4}_{3} \\
g^{i_1 i_1}_{4} & g^{i_2 j_2}_{4} & g^{i_3 j_3}_{4} & g^{i_4 j_4}_{4} 
\end{array}
\right|^+
= 2^3 \sum_{\pi \in \mathfrak{S}_4} \prod^4_{a=1} g^{i_a j_a}_{\pi (a)},
\end{align}
with the understanding that $j_1 \equiv i_1$. The remaining three raw coefficients are given by the same expression as eq. \eqref{eq:raw}, with the appropriate indices. To arrive at the final expression for the coefficient, the result of eq. \eqref{eq:spn_condensation} is that the raw coefficients of the permutations are scaled by $-\frac{1}{2}$,
\begin{align}
C^{(1)}_{i_1 j_2 j_3 j_4} &= 
D^{(1)}_{i_1 j_2 j_3 j_4} 
-\frac{1}{2} D^{(1)}_{j_2 i_1 j_3 j_4}
-\frac{1}{2} D^{(1)}_{j_3 j_2 i_1 j_4} 
-\frac{1}{2} D^{(1)}_{j_4 j_2 j_3 i_1} \\
&= 2^3 \sum_{\pi' \in \mathcal{S}_1} \sum_{\pi \in \mathfrak{S}_4} \text{sgn} (\pi' ) 
\prod^4_{a=1} g^{i_a j_{\pi' (a)}}_{\pi (a)}. \label{eq:eta_1_spn}
\end{align}
The set of permutations $\mathcal{S}_1 = \{ (1), (i_1 j_2), (i_1 j_3), (i_1 j_4) \}$ is \emph{not} a group, but we will clarify what it does represent in the next section.

Going to a CSF for $\eta=2$ introduces more terms, but no new complications, so we will proceed directly to 
\begin{align}
C^{(2)}_{i_1 i_2 j_3 j_4} &= D^{(2)}_{i_1 i_2 j_3 j_4} 
-\frac{1}{2} D^{(2)}_{i_2 i_1 j_3 j_4} -\frac{1}{2} D^{(2)}_{j_3 i_2 i_1 j_4} -\frac{1}{2} D^{(2)}_{j_4 i_2 j_3 i_1}
-\frac{1}{2} D^{(2)}_{i_1 j_3 i_2 j_4} -\frac{1}{2} D^{(2)}_{i_1 j_4 j_3 i_2} \nonumber \\
&+ \frac{1}{4} D^{(2)}_{j_3 i_1 i_2 j_4} +\frac{1}{4} D^{(2)}_{i_2 j_3 i_1 j_4} 
 + \frac{1}{4} D^{(2)}_{j_4 i_1 j_3 i_2} +\frac{1}{4} D^{(2)}_{i_2 j_4 j_3 i_1} 
 + \frac{1}{4} D^{(2)}_{j_3 j_4 i_1 i_2} +\frac{1}{4} D^{(2)}_{j_4 j_3 i_2 i_1} \\
&= 2^2 \sum_{\pi' \in \mathcal{S}_2} \sum_{\pi \in \mathfrak{S}_4} \text{sgn} (\pi' ) 
\prod^4_{a=1} g^{i_a j_{\pi' (a)}}_{\pi (a)},
\end{align}
where each condensation introduces a factor of $-\frac{1}{2}$. The set of summed permutations is 
\begin{align}
\mathcal{S}_2 =\{ &(1), (i_1 i_2), (i_1 j_3), (i_1 j_4), (i_2 j_3), (i_2 j_4), \nonumber \\
&(i_1 i_2 j_3), (i_1 j_3 i_2), (i_1 i_2 j_4), (i_1 j_4 i_2), (i_1 j_3)(i_2 j_4), (i_1 j_4)(i_2 j_3) \}.
\end{align}
The remaining expressions are
\begin{align}
C^{(3)}_{i_1 i_2 i_3 j_4} &= 2 \sum_{\pi' \in \mathcal{S}_3} \sum_{\pi \in \mathfrak{S}_4} \text{sgn} (\pi' ) 
\prod^4_{a=1} g^{i_a j_{\pi' (a)}}_{\pi (a)} \\
C^{(4)}_{i_1 i_2 i_3 j_4} &=   \sum_{\pi' \in \mathcal{S}_4} \sum_{\pi \in \mathfrak{S}_4} \text{sgn} (\pi' ) 
\prod^4_{a=1} g^{i_a j_{\pi' (a)}}_{\pi (a)},
\end{align}
where $\mathcal{S}_3$ and $\mathcal{S}_4$ are the complete set of $4!$ permutations. A suggestive pattern has emerged which will now be made precise.

\subsubsection{Expansion Coefficients: General Expressions}

Let us first consider the two extreme cases of $\eta = 0$ and $\eta = M$ before presenting the general result. For $\eta = 0$, a typical CSF in the expansion \eqref{eq:spN_state} wil look like
\begin{align}
A^+_{i_1 j_1} \dots A^+_{i_M j_M} \ket{\theta}.
\end{align}
Since all the indices $\{ (i,j) \}$ are distinct, no rearrangement of the indices will yield the same CSF. Therefore, for each $A^+_{ij}$ in the CSF, each geminal $G^{\dagger}_{\alpha}$ will contribute the factor $2g^{ij}_{\alpha}$ to the expansion coefficient. The result is therefore a symmetric sum, expressible as the permanent
\begin{align}
C^{(0)}_{ (i_1 j_1),\dots,(i_M j_M) } &= 2^M
	\left|
	\begin{array}{ccc}
		g^{i_1 j_1}_1 & & g^{i_M j_M}_1 \\
		& \ddots & \\
		g^{i_1 j_1}_M & & g^{i_M j_M}_M
	\end{array}
	\right|^+ \\
&= 2^M \sum_{\pi \in \mathfrak{S}_M} \prod^M_{a=1} g^{i_a j_a}_{\pi (a)}. \label{eq:spn_algebraic_permanent}
\end{align}
We have written the permanent algebraically, \eqref{eq:spn_algebraic_permanent} so that it can be easily compared with the closed-shell result $\eta = M$. In that case, there is not only a contribution from the term
\begin{align}
S^+_{i_1} \dots S^+_{i_M} \ket{\theta} \equiv A^+_{i_1 i_1} \dots A^+_{i_M i_M} \ket{\theta},
\end{align}
but also from each permutation $\pi' \in \mathfrak{S}_M$ of the second lower indices. From eq. \eqref{eq:spn_condensation}, each transposition gives a factor of $-\frac{1}{2}$, but each $A^+_{ij}$ occurs \emph{twice} in the geminal \eqref{eq:spN_geminal} so that the contribution is just the sign of the permutation. The expression for the expansion coefficient is therefore
\begin{align} \label{eq:coeff_md}
C^{(M)}_{(i_1 i_1),\dots, (i_M i_M)} = \sum_{\pi' \in \mathfrak{S}_M} \sum_{\pi \in \mathfrak{S}_M} \text{sgn} (\pi') \prod^M_{a=1} 
g^{i_a i_{\pi' (a)}}_{\pi (a)}, 
\end{align}
where there is a double summation over the entire symmetric group.

Now consider $0< \eta < M$, and arrange the list $\omega$ such that the distinct indices are at the end, i.e. the first $\eta$ pairs are identical, as in eq. \eqref{eq:omega_example}. The set of permutations of the last $M-\eta$ (open-shell) indices, along with the identity permutation, forms a subgroup $H \subset \mathfrak{S}_M$, which is equivalent to $H \cong \mathfrak{S}_{M - \eta}$. The action of each of these permutations will yield a different CSF.  We need to ``factor out'' the elements of $H$ from $\mathfrak{S}_M$ to arrive at the permutations which do not yield different CSFs. This is accomplished by looking at the left cosets of $H$, i.e. by constructing the sets $\pi H = \{\pi \pi' : \pi' \in H\}$, for each $\pi$. An elementary result from the theory of groups is Lagrange's theorem:
\begin{align}
|G| = [G:H] |H|
\end{align}
which states the order of a subgroup $H$ of $G$ divides the order of $G$, and the factor $[G:H]$ is the number of cosets. Each element of $G$ will occur in precisely one coset of $H$.
Returning to the example of the previous section, for the $\eta = 1$ case, 
\begin{align}
H &= \{ (1), (j_2 j_3), (j_2 j_4), (j_3 j_4), (j_2 j_3 j_4), (j_2 j_4 j_3)\}
\end{align}
and the corresponding left cosets are:
\begin{subequations}
\begin{align}
          H &= \{ (1), (j_2 j_3), (j_2 j_4), (j_3 j_4), (j_2 j_3 j_4), (j_2 j_4 j_3)\} \\
(i_1 j_2) H &= \{ (i_1 j_2), (i_1 j_2 j_3), (i_1 j_2 j_4), (i_1 j_2)(j_3 j_4), (i_1 j_2 j_3 j_4), (i_1 j_2 j_4 j_3)\} \\
(i_1 j_3) H &= \{ (i_1 j_3), (i_1 j_3 j_2), (i_1 j_3)(j_2 j_4), (i_1 j_3 j_4), (i_1 j_3 j_4 j_2), (i_1 j_3 j_2 j_4)\} \\
(i_1 j_4) H &= \{ (i_1 j_4), (i_1 j_4)(j_2 j_3), (i_1 j_4 j_2), (i_1 j_4 j_3), (i_1 j_4 j_2 j_3), (i_1 j_4 j_3 j_2)\}.
\end{align}
\end{subequations}
A quick glance will confirm that the coset representatives (those appearing on the left hand side) are precisely the elements of $\mathcal{S}_1$ to be summed over in eq. \eqref{eq:eta_1_spn}. Similarly, for $\eta =2$, we have
$H = \{ (1), (j_3 j_4)\}$, with the corresponding left cosets:

\begin{align*}
         H& = \{(i_1), (j_3,j_4)\}               &(i_1 i_2 j_3)H& = \{(i_1 i_2 j_3),(i_1 i_2 j_3 j_4)\} \\
(i_1 i_2)H& = \{(i_1 i_2),(i_1 i_2)(j_3 j_4)\}   &(i_1 j_3 j_2)H& = \{(i_1 j_3 i_2),(i_1 j_3 j_4 i_2)\} \\
(i_1 j_3)H& = \{(i_1 j_3),(i_1 j_3 j_4)\}        &(i_1 i_2 j_4)H& = \{(i_1 i_2 j_4),(i_1 i_2 j_4 j_3)\} \\
(i_1 j_4)H& = \{(i_1 j_4),(i_1 j_4 j_3)\}        &(i_1 j_4 i_2)H& = \{(i_1 j_4 i_2),(i_1 j_4 j_3 i_2)\} \\
(i_2 j_3)H& = \{(i_2 j_3),(i_2 j_3 j_4)\}   &(i_1 j_3)(i_2 j_4)H& = \{(i_1 j_3)(i_2 j_4),(i_1 j_3 i_2 j_4)\} \\
(i_2 j_4)H& = \{(i_2 j_4),(i_2 j_4 j_3)\}   &(i_1 j_4)(i_2 j_3)H& = \{(i_1 j_4)(i_2 j_3),(i_1 j_4 i_2 j_3)\},
\end{align*}
of which, the coset representatives are precisely $\mathcal{S}_2$. We are now in a position to write the general result for the expansion coefficient. Denote the coset representative of a specific coset $\pi H$ as $\natural (\pi H)$, e.g. $\natural ( (i_1 i_2) H) = (i_1 i_2)$. The expansion coefficient for a CSF with $0 \leq \eta \leq M$ and $\omega$ is
\begin{align} 
C^{(\eta)}_{\omega} = 2^{M-\eta} \sum_{ \{\pi' H\}} \sum_{\pi \in \mathfrak{S}_M} \text{sgn} (\natural (\pi' H))
\prod^M_{a=1} g^{i_a j_{ \natural (\pi' H)(a) }}_{\pi (a)}.
\end{align} 

\subsubsection{CSF Overlaps: Four Pair Example}

We will proceed in a similar manner to our approach for the expansion coefficients, beginning with the $\eta=4$ case as it is the easiest, and working down to $\eta=0$. CSFs with $\eta = 4$ are Slater determinants, and thus are orthonormal, i.e.
\begin{align}
\braket{\Phi^{\eta '}_{\omega '}  | \Phi^{(4)}_{\omega}} = \delta_{4,\eta'} \delta_{\omega \omega'}.
\end{align}

Next we consider the overlap $\braket{\Phi^{\eta '}_{\omega '}  | \Phi^{(3)}_{\omega}}$, with the arbitrary CSF
\begin{align}
\ket{\Phi^{(3)}_{\omega}} = S^+_{i_1} S^+_{i_2} S^+_{i_3} A^+_{ij} \ket{\theta}. 
\end{align}
First, the overlap can only be non-zero if $\eta' =3$, with identical indices for the $S^+$'s, i.e.
\begin{align}
\braket{\Phi^{(3)}_{\omega '}  | \Phi^{(3)}_{\omega}} &=
- \braket{\theta | A^-_{i'j'} S^-_{i_3'} S^-_{i_2'} S^-_{i_1'} S^+_{i_1} S^+_{i_2} S^+_{i_3} A^+_{ij} | \theta} \\
&= - \delta_{ \{ \eta\} \{\eta'\} } \braket{\theta | A^-_{i'j'} A^+_{ij} | \theta} \\
&= \delta_{ \{\eta\} \{\eta'\} } \braket{\theta | [A^+_{ij}, A^-_{i'j'} ] | \theta} \\
&= \frac{1}{2} \delta_{ \{\eta\} \{\eta'\} }
\left(  \delta_{ii'}\delta_{jj'} + \delta_{ij'}\delta_{ji'} \right).
\end{align}
Again, because the pair creator is symmetric, i.e. $A^+_{ij} = A^+_{ji}$, only one choice is possible. It is convenient to consider only \emph{bra} states with $i < j$, and therefore only the direct term contributes. The result is that
\begin{align}
\braket{\Phi^{(3)}_{\omega '}  | \Phi^{(3)}_{\omega}} &= \frac{1}{2} \delta_{\omega \omega'}.
\end{align}
The abbreviated Kronecker delta $\delta_{ \{\eta\} \{\eta'\} }$ denotes the equality of the two sets of indices of the first $\eta$ indices. It means that both $\eta = \eta'$, and the sets of indices are the same.

For $\eta=2$, again the indices for $S^+$'s must be identical, and using the sp(N) structure constants we arrive at 
\begin{align}
\braket{\Phi^{(2)}_{\omega'} | \Phi^{(2)}_{\omega}} &= (-1)^2 \delta_{ \{\eta\} \{\eta'\} } \braket{ \theta | A^-_{k'l'} A^-_{i'j'} A^+_{ij} A^+_{kl} |\theta }
\end{align}
with
\begin{align}
\braket{ \theta | A^-_{k'l'} A^-_{i'j'} A^+_{ij} A^+_{kl} |\theta }
&= \frac{1}{4} \left( \delta_{ii'}\delta_{jj'} + \delta_{ij'}\delta_{ji'} \right) \left( \delta_{kk'}\delta_{ll'} + \delta_{kl'}\delta_{lk'} \right) \nonumber \\
&+ \frac{1}{4} \left( \delta_{ik'}\delta_{jl'} + \delta_{il'}\delta_{jk'} \right) \left( \delta_{ki'}\delta_{lj'} + \delta_{kj'}\delta_{li'} \right) \nonumber \\
&- \frac{1}{8} \left( \delta_{ii'}\delta_{jk'} + \delta_{ij'}\delta_{ki'} \right) \left( \delta_{jk'}\delta_{ll'} + \delta_{jl'}\delta_{lk'} \right) \nonumber \\
&- \frac{1}{8} \left( \delta_{ii'}\delta_{lj'} + \delta_{ij'}\delta_{li'} \right) \left( \delta_{jk'}\delta_{kl'} + \delta_{jl'}\delta_{kk'} \right) \nonumber \\
&- \frac{1}{8} \left( \delta_{jj'}\delta_{ki'} + \delta_{ji'}\delta_{kj'} \right) \left( \delta_{ik'}\delta_{ll'} + \delta_{il'}\delta_{lk'} \right) \nonumber \\
&- \frac{1}{8} \left( \delta_{jj'}\delta_{li'} + \delta_{ji'}\delta_{lj'} \right) \left( \delta_{ik'}\delta_{kl'} + \delta_{il'}\delta_{kk'} \right).
\label{eq:4_2_big}
\end{align}
This expression is easily interpreted as an action of the symmetric group on the indices $\{i,j,k,l\}$, which is to say it is a function of the form
\begin{align}
F [i,j,k,l] = \sum_{\pi \in \mathfrak{S}_4} f (\pi) \delta_{\pi(i) i'}\delta_{\pi(j) j'}\delta_{\pi(k) k'}\delta_{\pi(l) l'},
\end{align}
and the result can be simplified substantially because of the property that the indices are ordered, and that the operators $A^+_{ij}$ commute with one another. The first two lines of eq. \eqref{eq:4_2_big} say that the CSF $\ket{A^+_{ij}A^+_{kl}}$ has overlap $\frac{1}{4}$ with the CSFs $\ket{A^+_{ij}A^+_{kl}}$, 
$\ket{A^+_{ij}A^+_{lk}}$, $\ket{A^+_{ji}A^+_{kl}}$, $\ket{A^+_{ji}A^+_{lk}}$, $\ket{A^+_{kl}A^+_{ij}}$, $\ket{A^+_{lk}A^+_{ij}}$, $\ket{A^+_{kl}A^+_{ji}}$, and 
$\ket{A^+_{lk}A^+_{ji}}$, \emph{but}, these CSFs are identical, and we only consider $\bra{A^+_{ij}A^+_{kl}}$ in our set of states to project against. The overlap is then substantially simplified
\begin{align}
\braket{\Phi^{(2)}_{\omega'} | \Phi^{(2)}_{\omega}} &= \frac{1}{4} \delta_{\{\eta\}\{\eta'\}}
\left(
             \delta_{ii'}\delta_{jj'}\delta_{kk'}\delta_{ll'} 
-\frac{1}{2} \delta_{ii'}\delta_{kj'}\delta_{jk'}\delta_{ll'} 
-\frac{1}{2} \delta_{ii'}\delta_{lj'}\delta_{kk'}\delta_{jl'} 
\right).
\end{align}
As in the previous section, this argument may be formalized with cosets. We are dealing with the symmetric group on the four letters $i,j,k,l$. The set of eight permutations of indices which leave the CSF $\ket{A^+_{ij}A^+_{kl}}$ invariant
\begin{align}
H &= \{ (1), (ij), (kl), (ij)(kl), (ik)(jl), (il)(jk), (ikjl), (iljk) \}
\end{align}
close a subgroup of $\mathfrak{S}_4$. To arrive at the distinct CSFs, we again ``factor out'' the action of the subgroup by looking at (right) cosets.\footnote{This subgroup is in general not \emph{normal} as it is well known that for $N>4$ the only normal subgroup of $\mathfrak{S}_N$ is $A_N$, the alternating group on $N$ objects. Thus the left and right cosets are not the same, and the resulting object is not properly a factor group. Nonetheless, we use the term ``factoring'' but we will continue to be precise by labelling cosets appropriately.} By Lagrange's theorem, there are three distinct cosets, which are:
\begin{align}
    H &= \{  (1), (ij), (kl), (ij)(kl), (ik)(jl), (il)(jk), (ikjl), (iljk) \} \\
H(jk) &= \{ (il), (jk), (ijk), (ikl), (ilj), (jlk), (ijlk), (iklj) \} \\
H(jl) &= \{ (ik), (jl), (ijl), (ikj), (ilk), (jkl), (ijkl), (ilkj) \}.
\end{align}
The coset labels are arbitrary, in that each coset member is as good a label as any other. What does matter is the minimum number of transpositions, which in both cases is 1. We can clean up the expression to include a summation over right cosets:
\begin{align}
\braket{\Phi^{(2)}_{\omega'} | \Phi^{(2)}_{\omega}} &= \frac{1}{4} \delta_{\{\eta\}\{\eta'\}}
\sum_{\{H \pi\}} \left( - \frac{1}{2} \right)^{ \flat(H\pi) }
\delta_{\pi(i)i'}\delta_{\pi(j)j'}\delta_{\pi(k)k'}\delta_{\pi(l)l'} 
\end{align}
The number $\flat (H\pi)$ is the \emph{minimum} number of transpositions required to write the coset representative of $H\pi$. The number of transpositions is not a specific number, but the minimum number is. In this case, $\flat ( H ) = 0$, while those of the other two cosets is $\flat ( H (jk) ) = \flat (H(jl))=1$. 

Passing to $\eta =1$ introduces no algebraic complication, though the complete expression is intractable to report. We must deal with $\mathfrak{S}_6$, which has $6! = 720$ elements. As was the case previously, several permutations will leave a CSF invariant, and we will factor these out (i.e., consider only right cosets). Specifically, $2^3 \cdot 6 = 48$ permutations leave the CSF $\ket{A^+_{ij}A^+_{kl}A^+_{mn}}$ invariant, and we can deduce this number combinatorially. First, the order of the indices $ij$ in $A^+_{ij}$ do not matter, so each $A^+$ produces a factor of 2, hence $2^3$. Second, the order of the 3 $A^+$'s does not matter, since they commute, so there is a factor of $3! = 6$. By Lagrange's theorem, there are $\frac{720}{48}=15$ cosets. The expression for the overlap is quite similar to the previous case, 
\begin{align}
\braket{\Phi^{(1)}_{\omega'} | \Phi^{(1)}_{\omega}} &= \frac{1}{8} \delta_{\{\eta\}\{\eta'\}}
\sum_{\{H \pi\}} \left( - \frac{1}{2} \right)^{ \flat(H\pi) }
\delta_{\pi(i)i'}\delta_{\pi(j)j'}\delta_{\pi(k)k'}\delta_{\pi(l)l'} \delta_{\pi(m)m'}\delta_{\pi(n)n'}
\end{align}
with the distinction being in the members of the sum $H\pi$. The subgroup $H$ itself has weight 1, while the cosets with labels $(jk), (jl), (jm), (jn), (lm), (ln)$ have weight $-\frac{1}{2}$, and those with labels $(jk)(lm), (jk)(ln), (jl)(km), (jl)(kn), (jkm), (jkn), (jlm), (jln)$ have weight $+\frac{1}{4}$. A definite pattern is now apparent. Without writing the explicit result, the $\eta = 0$ case will have a sum over $\frac{8!}{2^4\cdot 4!}=105$ right cosets.

\subsubsection{CSF Overlaps: General Expressions}
Based on the discussion of the previous section, the overlap between singlet CSFs of $2M$ electrons with $\eta$ pairs is:
\begin{align}
\braket{\Phi^{(\eta')}_{\omega'} | \Phi^{(\eta)}_{\omega}} &=
\frac{1}{2^{(M-\eta)}} \delta_{\{\eta\} \{\eta'\}} 
\sum_{\{H \pi\}} \left( - \frac{1}{2} \right)^{ \flat(H\pi) }
\prod^{2(M-\eta)}_{a=1} \delta_{\pi (i_a) i_a'}.
\end{align}
The summation is performed over right cosets of the group of permutations of $\mathfrak{S}_{2(M-\eta)}$. Each coset contains $2^{(M-\eta)}(M-\eta)!$ elements: for the product of $(M-\eta)$ objects $A^+_{ij}$, the order of the indices doesn't matter so each contributes a factor 2. Further, the order of the $A^+_{ij}$ doesn't matter, hence the factor $(M-\eta)!$. By Lagrange's theorem, the number of cosets is:
\begin{align}
\frac{(2(M-\eta))!}{2^{(M-\eta)}(M-\eta)!} = (2(M-\eta)-1)!!
\end{align}
here the double factorial of an integer is
\begin{align}
n!! = n (n-2)(n-4)...
\end{align}
Again, the number $\flat (H\pi)$ is the minimum number of transpositions which reproduce the coset representative of $H\pi$.

\subsection{Discussion}
Now that the expansion coefficients and CSF overlaps are symbolically understood, a brief discussion of feasibility is warranted. The expansion coefficient \eqref{eq:coeff_md} is known as a mixed discriminant:\cite{md1,md2,md3} a complete double summation over the entire symmetric group, one symmetric, the other antisymmetric. The computation of a mixed discriminant is not feasible in general, so special cases must be considered to move forward. An obvious avenue is to consider a geminal coefficient 
\begin{align}
	g^{ij}_{\alpha} = f^i_{\alpha} f^j_{\alpha},
\end{align}
though this causes the sp(N) geminal to factorize
\begin{align}
	G^{\dagger}_{\alpha} = \sum_{ij} f^i_{\alpha} f^j_{\alpha} A^+_{ij} = \alpha^{\dagger}_{\uparrow} \alpha^{\dagger}_{\downarrow}
\end{align}
into orbitals
\begin{align}
	\alpha^{\dagger}_{\sigma} = \sum_i f^i_{\alpha} a^{\dagger}_{i\sigma}.
\end{align}
The resulting geminal product is a Slater determinant of doubly occupied orbitals.

For su(2) geminals, i.e. APIG, the expansion coefficients are permanents which are not feasible either. However, APIG becomes feasible in three different ways.\cite{moisset:2022a} First, if all the geminal coefficients are the same, $g^{ij}_{\alpha} = g^{ij}$, then the mixed discriminant reduces to a determinant weighted by the factor $M!$
\begin{align}
	C^{(M)}_{(i_1 i_1),\dots, (i_M i_M)} &= M! \begin{vmatrix}
		g^{i_1 i_1} &  & g^{i_1 i_M} \\
		& \ddots & \\
		g^{i_M i_1} &  & g^{i_M i_M}
		\end{vmatrix},
\end{align}
though the geminal coefficients can be diagonalized and the geminal product state reduces to the antisymmetrized geminal power (AGP) in its natural orbitals, an su(2)-geminal wavefunction. This \emph{may} be a manner to remove the orbital-optimization from AGP. 

Next, the geminal subspaces can be fixed so that each orbital belongs to only one. The corresponding geminals are \emph{strongly orthogonal} and the resulting geminal product is known as APSG. In this case, many of the geminal coefficients are zero, there are only a small number of mixed discriminants to compute, and the mixed discriminants themselves are much simpler. 

Similarly, one could restrict the geminal coefficients such that for $k=1,\dots,\frac{N}{2}$ the only non-zero possibilities are $g^{(2k-1)(2k-1)}_{\alpha}$, $g^{(2k-1)(2k)}_{\alpha}=g^{(2k)(2k-1)}_{\alpha}$ and $g^{(2k)(2k)}_{\alpha}$. This amounts to using a collection of so(5) copies rather than sp(N). Further structure of the geminal coefficients would still be necessary to arrive at tractable expressions. A related construction is 2D-block geminals, when restricted to singlets.\cite{cassam:2023} This construction also employs so(5) copies, but in addition: each so(5) copy contributes closed-shell pairs to one geminal and open-shell singlet pairs to a \emph{different} geminal. For example, the $k$th so(5) copy has geminal coefficients $g^{(2k-1)(2k-1)}_{\alpha}$ and $g^{(2k)(2k)}_{\alpha}$ in the $\alpha$th geminal, and $g^{(2k-1)(2k)}_{\alpha} = g^{(2k)(2k-1)}_{\alpha} = 0$, but also in the $\beta$th geminal $g^{(2k-1)(2k-1)}_{\beta} = g^{(2k)(2k)}_{\beta} =0$ and $g^{(2k-1)(2k)}_{\beta} = g^{(2k)(2k-1)}_{\beta} \neq 0$. For any other geminal $\gamma$, there are no non-zero coefficients from the $k$th so(5) copy $g^{(2k-1)(2k-1)}_{\gamma} = g^{(2k)(2k)}_{\gamma} =g^{(2k-1)(2k)}_{\gamma} = g^{(2k)(2k-1)}_{\gamma} = 0$.

There may be structured geminal coefficients that lead to mixed discriminants that are computable as a small number of determinants. The strongest known result for permanents is due to Carlitz and Levine:\cite{carlitz:1960} for the matrix $P$ with elements
\begin{align}
	P = \begin{pmatrix}
		p_{11} & & p_{1N} \\
		& \ddots & \\
		p_{N1} & & p_{NN}
	\end{pmatrix}
\end{align}
the permanent of the matrix with elements $p^{-1}_{ij}$ times its determinant is the determinant of the matrix with elements $p^{-2}_{ij}$
\begin{align} \label{eq:carlitz_levine}
	\begin{vmatrix}
		p^{-1}_{11} & & p^{-1}_{1N} \\
		& \ddots & \\
		p^{-1}_{N1} & & p^{-1}_{NN}
	\end{vmatrix}^+
	\begin{vmatrix}
		p^{-1}_{11} & & p^{-1}_{1N} \\
		& \ddots & \\
		p^{-1}_{N1} & & p^{-1}_{NN}
	\end{vmatrix}
	= 
	\begin{vmatrix}
		p^{-2}_{11} & & p^{-2}_{1N} \\
		& \ddots & \\
		p^{-2}_{N1} & & p^{-2}_{NN}
	\end{vmatrix},
\end{align}
provided that all 3 $\times$ 3 minors of $P$ vanish ($P$ has rank at most 2). A sufficient condition for this property is that the elements $p_{ij}$ satisfy Pl\"{u}cker conditions
\begin{align} \label{eq:plucker}
	p_{ij} p_{kl} - p_{ik} p_{jl} + p_{il} p_{jk} = 0.
\end{align}
These types of conditions arise naturally when embedding a small vector space into a larger one. Different applications have been identified previously in quantum chemistry.\cite{cassam:1992,cassam:1996} For any complex numbers $\{z\}$, the choices
\begin{align}
	p_{ij} &= z_i - z_j \\
	p_{ij} &= \exp(z_i - z_j) - \exp(-(z_i - z_j))
\end{align}
are solutions of \eqref{eq:plucker}. The first choice leads to geminal coefficients that are rational functions, \eqref{eq:carlitz_levine} is Borchardt's theorem,\cite{borchardt} and the geminal products have the structure of RG states. The second choice leads to geminal coefficients that are trigonometric or hyperbolic functions and the geminal products have the structure of anisotropic RG states. At present there are no known mixed discriminants that simplify in such a manner. It is possible to construct RG states for sp(N), though at present it is not known how to reduced them to feasible expressions. There is exploration to be done.

Finally, sp(N) geminals should be obtainable as particular cases of known diagrammatic\cite{paldus:1972a,paldus:1972b} and algebraic\cite{cassam:2023} results for general geminal products. In particular, it is equivalent to GSCF presented in refs.\citenum{cassam:2007} and \citenum{cassam:2010}, though GSCF is solved iteratively in quite a different manner.

Thus, in general sp(N) geminals are unfeasible though there are options to consider. To demonstrate that it is worth pursuing this approach, we will now look at small strongly correlated systems numerically.

\section{Numerical Examples} \label{sec:results}
Variational calculations were performed for a set of 4-electron systems to demonstrate the benefit of adding open-shell singlets to the geminal structure. A 2-pair wavefunction built from sp(4)
\begin{align}\label{eq:spn_var}
	\ket{\textbf{sp(4)},2} &= \left(\sum_{ij} g^{ij}_1  A^{+}_{ij} \right) \left(\sum_{kl}g^{kl}_2 A^+_{kl}\right) \ket{\theta} \\
	&= \sum_{i<j} C_{iijj} S^+_i S^+_j \ket{\theta}
	+ \sum_i \sum_{j<k (\neq i)} C_{iijk} S^+_i A^+_{jk} \ket{\theta} \nonumber \\
	&+ \sum_{i<j} \sum_{k<l (\neq i,j)} C_{ijkl} A^+_{ij} A^+_{kl} \ket{\theta}
\end{align}
is defined in terms of the expansion coefficients
\begin{align}
	C_{iijj} &= g^{ii}_1 g^{jj}_2 + g^{jj}_1 g^{ii}_2 - 2 g^{ij}_1 g^{ij}_2 \\
	C_{iijk} &= 2 \left( g^{ii}_1 g^{jk}_2 + g^{jk}_1 g^{ii}_2 - g^{ij}_1 g^{ik}_2 - g^{ik}_1 g^{ij}_2 \right) \\
	C_{ijkl} &= 4 \left( g^{ij}_1 g^{kl}_2 + g^{kl}_1 g^{ij}_2 \right).
\end{align}
With the implied symmetries $C_{iijj} = C_{jjii}$, along with 
\begin{align}
	C_{iijk} &= C_{iikj} = C_{jkii} = C_{kjii} \\
	C_{ijkl} &= C_{ijlk} = C_{jikl} = C_{jilk} = C_{klij} = C_{lkij} = C_{klji} = C_{lkji}
\end{align}
the expression for the norm is
\begin{align}
	\braket{\textbf{sp(4)},2 | \textbf{sp(4)},2} &=
	\sum_{i<j} C_{iijj} C_{iijj} + \frac{1}{2} \sum_i \sum_{j<k (\neq i)} C_{iijk} C_{iijk} \nonumber \\
	&+ \frac{1}{8} \sum_{i<j} \sum_{k<l (\neq i,j)} C_{ijkl} \left( C_{ijkl} - \frac{1}{2} C_{ikjl} - \frac{1}{2} C_{iljk} \right).
\end{align}
This last bracketed term arises from the non-orthogonality of the CSF basis, a direct result of the braiding \eqref{eq:braiding}. The energy expression to be minimized is a function of the geminal coefficients $\{g\}$
\begin{align} \label{eq:obj}
E [\{g\}] &= \frac{\braket{\textbf{sp(4)},2|\hat{H}|\textbf{sp(4)},2}}{\braket{\textbf{sp(4)},2 | \textbf{sp(4)},2}} \\
&= \sum_{ij} h_{ij} d_{ij} + \frac{1}{2} \sum_{ijkl} V_{ijkl} d_{ijkl} + V_{NN}
\end{align}
in terms of the 1- and 2-body reduced density matrix elements
\begin{align}
	d_{ij}   &= - 2 \braket{\textbf{sp(4)},2|A^0_{ij}|\textbf{sp(4)},2} \\
	d_{pqrs} &=     \braket{\textbf{sp(4)},2|4 A^0_{ij} A^0_{kl} + 2 \delta_{jk}A^0_{il}|\textbf{sp(4)},2}.
\end{align}
Explicit expressions for unnormalized $d_{ij}$ and $d_{ijkl}$ are included in appendix \ref{sec:rdm}. They are straightforward to compute. As these computations are brute-force there is no reason to be clever: the objective function \eqref{eq:obj} was minimized with a combination of the covariance matrix adaptation evolution strategy (CMA-ES)\cite{hansen:2001} and the Nelder-Mead\cite{nelder:1965} simplex algorithm.

The first strongly correlated system of electrons we consider is linear equidistant H$_4$ in a minimal STO-6G basis. 
\begin{figure}
	\begin{subfigure}{\textwidth}
		\includegraphics[width=0.485\textwidth]{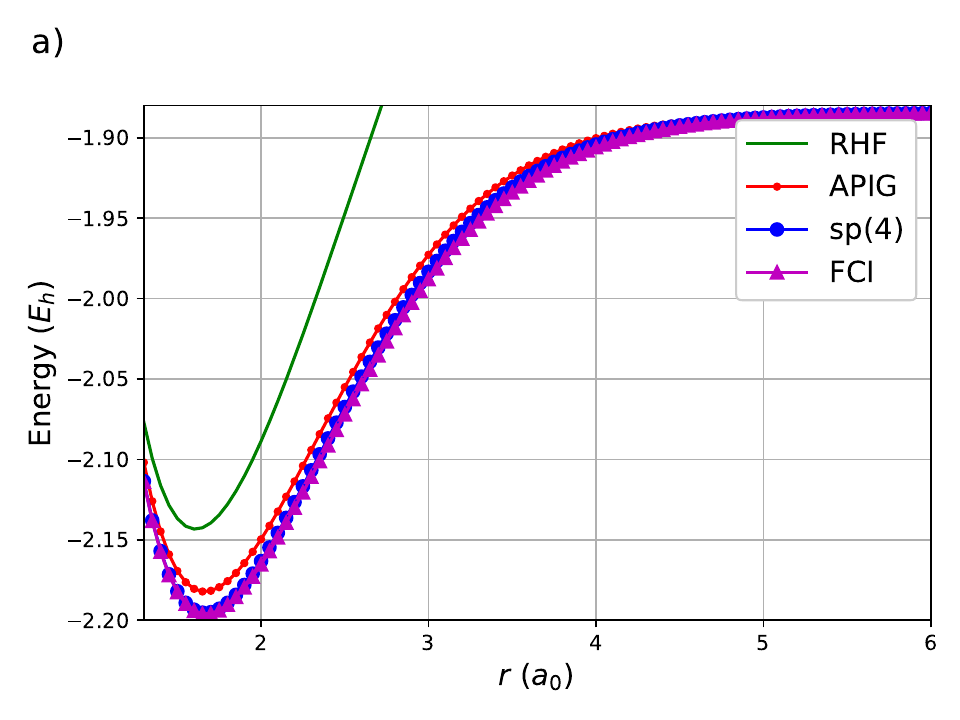}  \hfill
		\includegraphics[width=0.485\textwidth]{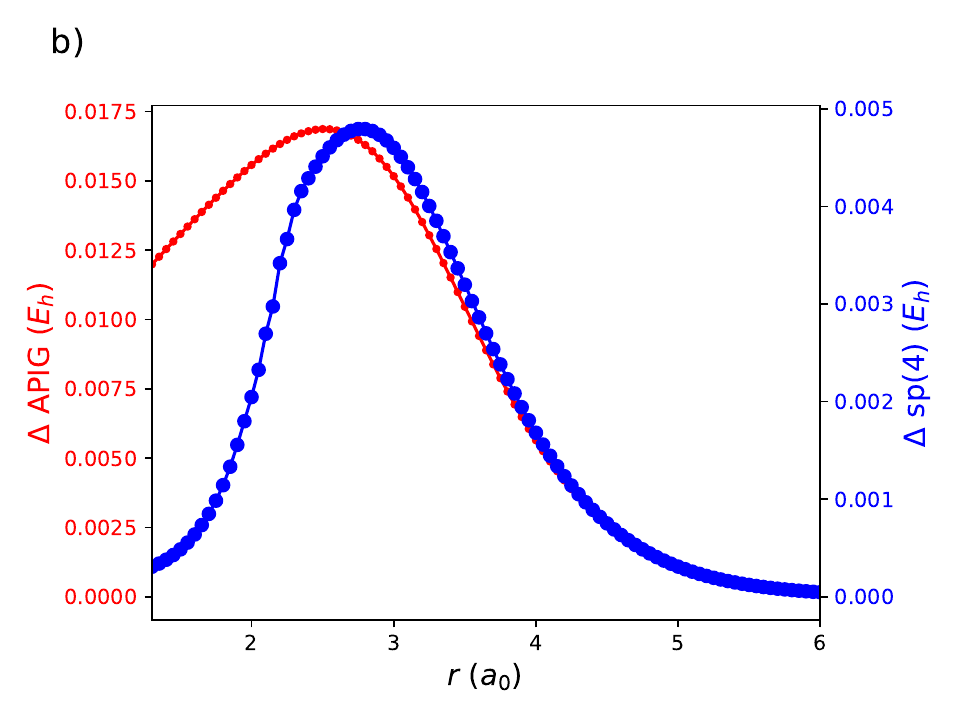}				
	\end{subfigure}
	\caption{Variational treatment of equidistant H$_4$ chain in the STO-6G basis: (a) RHF, APIG, sp(4), and FCI energies. (b) Error of sp(4) and APIG with respect to FCI. APIG results are computed in the basis of OO-DOCI orbitals.\cite{johnson:2020}}
	\label{fig:lin_H4_curves}
\end{figure}
Variational sp(4) curves are shown in figure \ref{fig:lin_H4_curves}, along with restricted Hartree-Fock (RHF), APIG, and full configuration interaction (FCI). RHF results were computed with Gaussian 16,\cite{gaussian_16} and FCI results were computed with psi4,\cite{sherill:1999,parrish:2017} both for ref. \citenum{johnson:2020}. APIG results were computed from brute-force expressions presented in ref. \citenum{moisset:2022a}. APIG is not invariant to orbital rotation, thus APIG was computed in the orbital-optimized (OO-)DOCI orbitals which were computed with GAMESS (US),\cite{barca:2020} also for ref. \citenum{johnson:2020}. OO-DOCI is the best possible result one could achieve with closed-shell singlet geminals. Strictly speaking, APIG is a variational approximation to OO-DOCI, though for all systems studied they are virtually identical. The APIG and sp(4) curves are visually discernible in figure \ref{fig:lin_H4_curves} (a) indicating a clear imporovement over APIG using sp(4). However, from figure \ref{fig:lin_H4_curves} (b) it is clear that sp(4) still misses 5mE$_h$ near a separation of $r=3a_0$. The OO-DOCI orbitals for linear hydrogen chains are simple: they form bonding/antibonding pairs on adjacent hydrogens.\cite{ward_thesis} There is one pair of bonding/antibonding orbitals on hydrogens 1 and 2, and one pair of bonding/antibonding orbitals on hydrogens 3 and 4. We will call such an arrangement of bonding/antibonding orbitals a \emph{pairing scheme}. APIG is a very good first approximation, missing only effects of weak correlation from open-shell excitations which sp(4) can include. The RHF curve is included mainly to show how poor it is. In the other studied systems it will be as bad or worse, and as such will be omitted.

The other systems are those considered by Paldus and coworkers,\cite{paldus:1993} which are colloquially known as the Paldus isomers of H$_4$. These systems have also recently been studied in a related context.\cite{gaikwad:2024} FCI results for these systems were computed using psi4\cite{smith:2020} in ref. \citenum{johnson:2023}. OO-DOCI results were also computed in ref. \citenum{johnson:2023} and the orbitals are used to compute APIG. The first system, called S4 in ref. \citenum{paldus:1993}, is a square of 4 H atoms with a constant side length $\alpha$. 
\begin{figure}
	\begin{subfigure}{\textwidth}
		\includegraphics[width=0.485\textwidth]{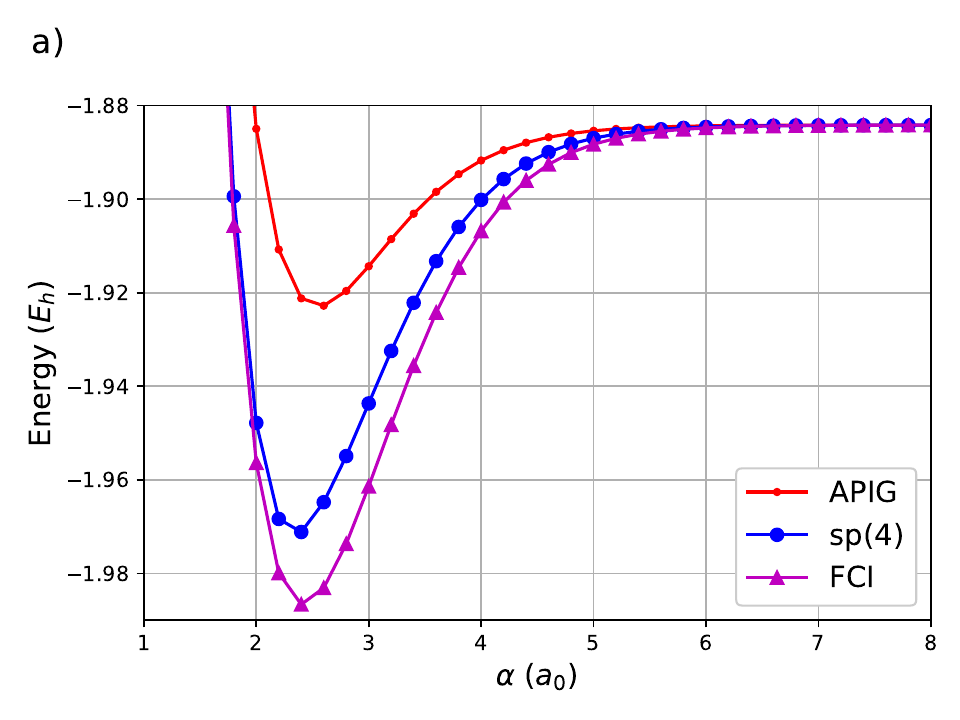}  \hfill
		\includegraphics[width=0.485\textwidth]{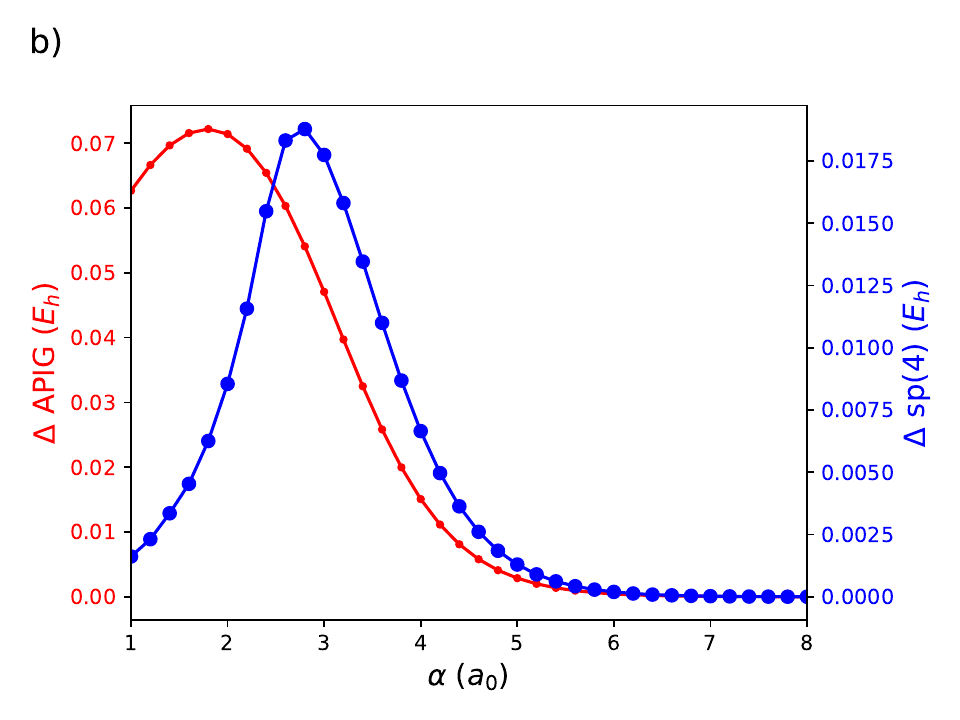}				
	\end{subfigure}
	\caption{Variational treatment of Paldus S4 (equidistant square H$_4$) in the STO-6G basis: (a) sp(4), APIG, and FCI energies. (b) Error of sp(4) and APIG with respect to FCI. APIG results are computed in the basis of OO-DOCI orbitals.\cite{johnson:2023}}
	\label{fig:S4_curves}
\end{figure}
FCI along with variational APIG and sp(4) curves are shown in figure \ref{fig:S4_curves}. One can see immediately that APIG is much worse than for linear H$_4$. Here, there are two degenerate pairing schemes: bonding/antibonding pairs can form in two ways. The correct answer includes a contribution from both, but APIG only accounts for one. In addition, there is weak correlation from open-shell excitations from both pairing schemes which APIG neglects entirely. Variational sp(4) is better than APIG but is still quite far from FCI. The sp(4) geminal can account for the degenerate pairing schemes, but still misses effects of weak correlation. The shapes of the errors, in figure \ref{fig:S4_curves} (b) are similar to those for linear H$_4$.

The next system, H4 in ref. \citenum{paldus:1993}, is a ring-opening of square H$_4$ to linear H$_4$. The H--H distances are fixed throughout the process, and 3 different lengths are considered: $a=1.2a_0$, $a=1.6a_0$, and $a=2.0a_0$. The length $a=1.6a_0$ is close to the equilibrium value, while $a=1.2a_0$ is compressed and $a=2.0a_0$ is lengthened. Effects of strong correlation increase as $a$ is increased: the bond weakens and the orbitals are closer to degenerate. In terms of the variable $\alpha$, the interior angle increases from $\pi \left(\alpha + \frac{1}{2}\right) = \frac{\pi}{2}$  to  $\pi \left(\alpha + \frac{1}{2}\right) = \pi$ describing the transition from square ($\alpha =0$) to linear ($\alpha = \frac{1}{2}$). 
\begin{figure}
	\begin{subfigure}{\textwidth}
		\includegraphics[width=0.485\textwidth]{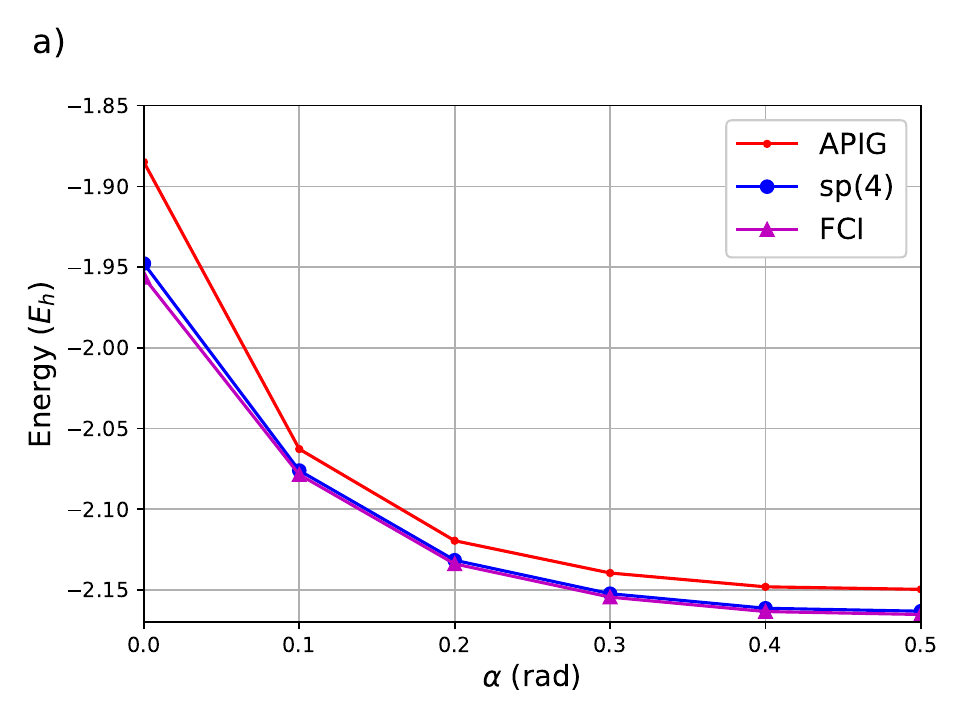}  \hfill
		\includegraphics[width=0.485\textwidth]{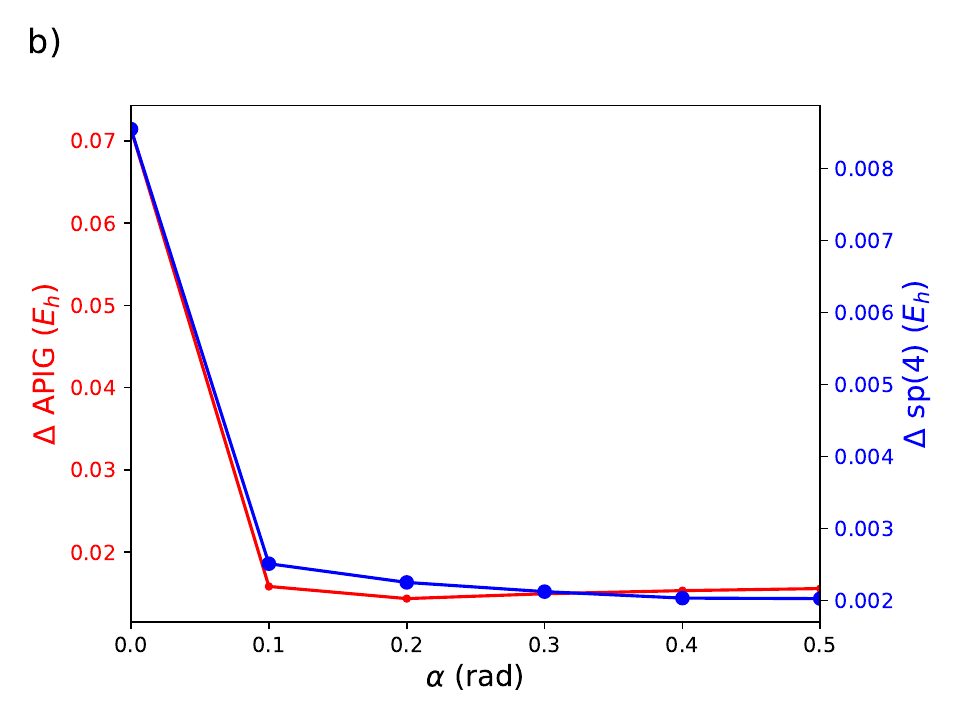}				
	\end{subfigure}
	\caption{Variational treatment of Paldus H4 (ring-opening of square to linear H$_4$) for $a=2.0a_0$ in the STO-6G basis: (a) sp(4), APIG, and FCI energies. (b) Error of sp(4) and APIG with respect to FCI. APIG results are computed in the basis of OO-DOCI orbitals.\cite{johnson:2023}}
	\label{fig:open_H4_curves}
\end{figure}
FCI, APIG, and sp(4) curves are shown for $a=2.0$ in figure \ref{fig:open_H4_curves}. Both APIG and sp(4) perform poorly at the square geometry ($\alpha=0$) due to the two degenerate pairing schemes. However, once the ring is opened the degeneracy is broken so that both APIG and sp(4) are essentially parallel to FCI, with APIG being one order of magnitude worse than sp(4). Similar results for $a=1.2a_0$ and $a=1.6a_0$ are shown in figures \ref{fig:open_H4_curves_1p2} and \ref{fig:open_H4_curves_1p6} in appendix \ref{sec:paldus}. The error for APIG is more or less the same for all three distances $a$ whereas sp(4) improves as $a$ is decreased.

D4 is a linear system with a fixed length $a$ between the first and second H atoms, and the third and fourth H atoms. The distance $\alpha$ between the second and third H atoms is varied. The same three values for $a$ are again studied.
\begin{figure}
	\begin{subfigure}{\textwidth}
		\includegraphics[width=0.485\textwidth]{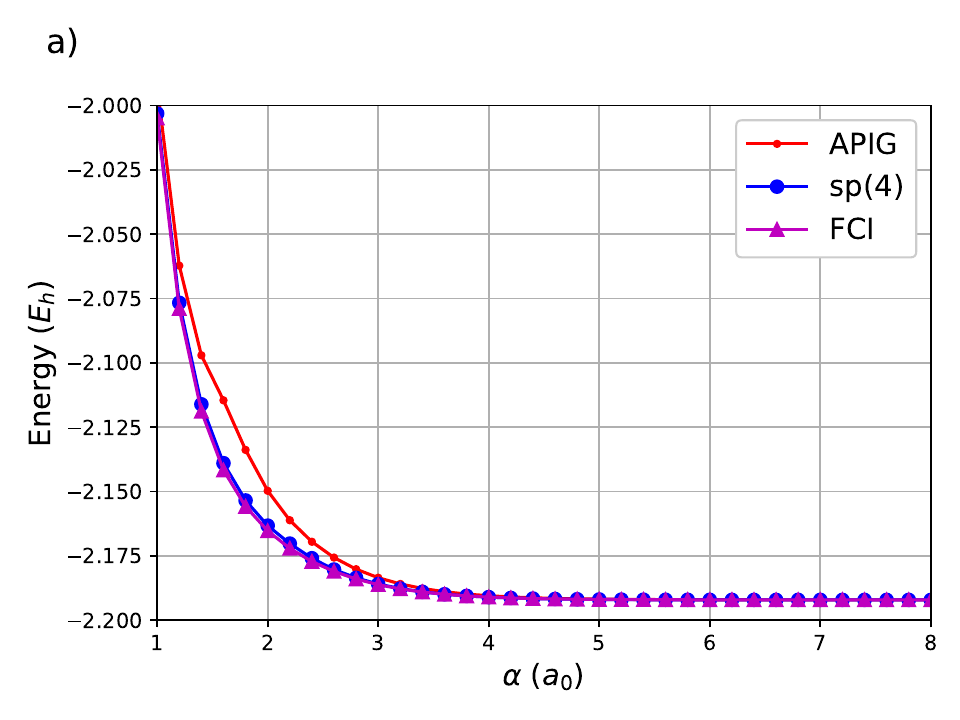}  \hfill
		\includegraphics[width=0.485\textwidth]{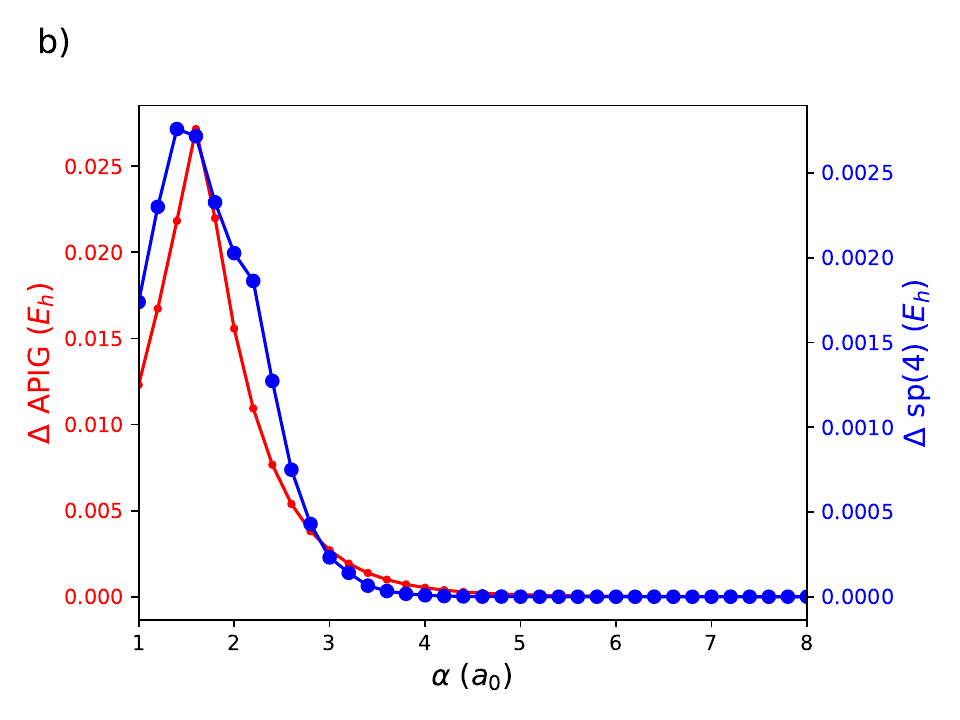}				
	\end{subfigure}
	\caption{Variational treatment of Paldus D4 for $a=2.0a_0$ in the STO-6G basis: (a) sp(4), APIG, and FCI energies. (b) Error of sp(4) and APIG with respect to FCI. APIG results are computed in the basis of OO-DOCI orbitals.\cite{johnson:2023}}
	\label{fig:D4_curves}
\end{figure}
One can see from the sp(4) results for $a=2.0a_0$ in figure \ref{fig:D4_curves} that there is difficulty when $\alpha < a$. In this region, the second and third H atoms are closer to one another than the terminal H atoms, meaning that it is better described as H--(H$_2$)--H than (H$_2$)--(H$_2$). The APIG results are again worse than sp(4), and the shapes of the errors are again similar. Here APIG is an entire order of magnitude worse than sp(4). Results for $a=1.2a_0$ and $a=1.6a_0$ are shown in figures \ref{fig:D4_curves_1p2} and \ref{fig:D4_curves_1p6} in appendix \ref{sec:paldus}, where APIG is much worse than sp(4).

Finally, P4 is a rectangular arrangement with a fixed side-length $a$ and a variable length $\alpha$. The same three particular choices of $a$ are considered. 
\begin{figure}
	\begin{subfigure}{\textwidth}
		\includegraphics[width=0.485\textwidth]{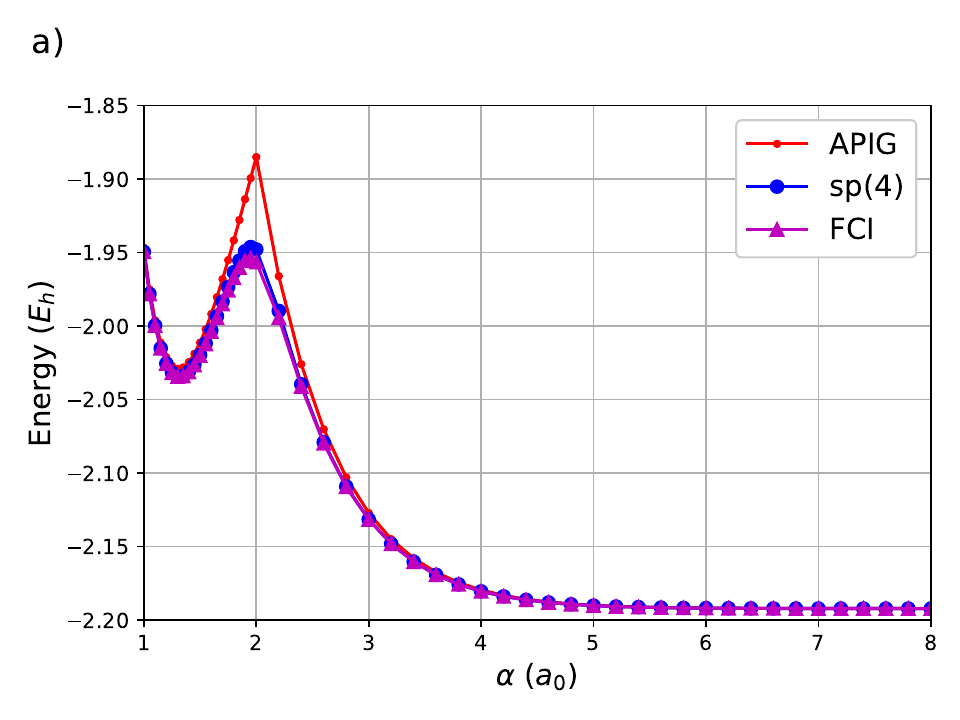}  \hfill
		\includegraphics[width=0.485\textwidth]{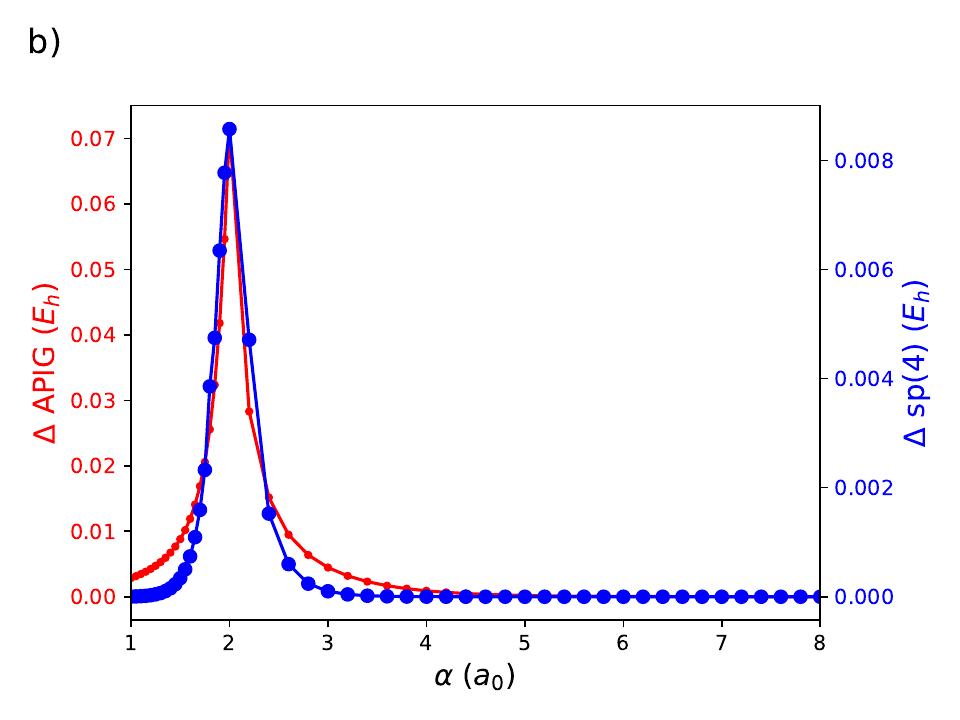}				
	\end{subfigure}
	\caption{Variational treatment of Paldus P4 for $a=2.0a_0$ in the STO-6G basis: (a) sp(4), APIG, and FCI energies. (b) Error of sp(4) and APIG with respect to FCI. APIG results are computed in the basis of OO-DOCI orbitals.\cite{johnson:2023}}
	\label{fig:P4_curves}
\end{figure}
FCI, APIG, and sp(4) curves are shown for $a=2.0a_0$ in figure \ref{fig:P4_curves}. There are two possible pairing schemes, one centred on the fixed length $a$ and the other on the variable length $\alpha$, with the more stable choice being the one centred on the shorter length. Again, at the square geometry $a=\alpha$ the schemes are degenerate. As a result APIG has a curve crossing at $a=\alpha$ whereas sp(4) and FCI have avoided crossings. The APIG error is again one order of magnitude worse than the sp(4) error. Similar results for $a=1.2a_0$ and $a=1.6a_0$ are shown in figures \ref{fig:P4_curves_1p2} and \ref{fig:P4_curves_1p6} in appendix \ref{sec:paldus}.

For all of these systems, variational sp(4) is a large improvement over APIG and a good approximation to FCI in a minimal basis. Thus, a pSE approach should be expected to perform well and approximations could be developed to make such an approach feasible. However, even in these small systems, sp(4) is missing effects of weak correlation which will get worse in larger systems. It will be very difficult to account for these effects. The usual treatment of the Random Phase Approximation has been shown to not give great results starting from closed-shell singlet wavefunctions,\cite{vu:2019} and it is difficult to converge. Likewise, perturbation theories are difficult to define without the low-lying excitations of the reference state, here sp(4). We should thus consider a model Hamiltonian, like the reduced BCS Hamiltonian, whose eigenvectors are sp(N) geminals. This is a possible way forward.\cite{holdhusen:2021}

\section{Conclusion}
Geminal wavefunctions for singlets have been constructed from the Lie algebra sp(N), yielding the ingredients required for a projected Schr\"{o}dinger equation: Hamiltonian action on basis vectors, expansion coefficients, and basis vector overlaps. The formulae are in general intractable so approximations will be required. Numerical results were obtained variationally for a collection of strongly correlated four electron systems. In each case, substantial improvement was found over closed-shell singlet pairs. The remaining effects of weak correlation will need to be included. 

\section{Acknowledgements}
PAJ and PWA acknowledge support from NSERC. This research was enabled in part by the Digital Research Alliance of Canada.

\appendix
\section{Reduced Density Matrix Elements for 4-site 2-pair singlets} \label{sec:rdm}
The 1-body reduced density matrix has diagonal
\begin{align}
	d_{aa} &= 2 \sum_{i (\neq a)} C_{aaii} C_{aaii} 
	+ \sum_{i < j (\neq a)} \left( C_{aaij} C_{aaij} + \frac{1}{2} C_{iiaj}C_{iiaj} + \frac{1}{2} C_{jjai}C_{jjai} \right) \nonumber \\
	&+ \frac{1}{4} \sum_{i<j<k (\neq a)} \left( C_{aijk}C_{aijk} + C_{ajik}C_{ajik} + C_{akij}C_{akij} \right) \nonumber \\
	&- \frac{1}{4} \sum_{i<j<k (\neq a)} \left( C_{aijk}C_{ajik} + C_{aijk}C_{akij} + C_{ajik}C_{akij} \right)
\end{align}
and off-diagonal elements.
\begin{align}
	d_{ab} &= d_{ba} \\
	 &= \sum_{i (\neq a,b)} \left( C_{iiab} \left( C_{aaii} + C_{bbii} \right) - \frac{1}{2} C_{aabi}C_{bbai} \right) \nonumber \\
	&+ \frac{1}{2} \sum_{i<j (\neq a,b)} \left( C_{iiaj}C_{iibj} + C_{jjai}C_{jjbi} + \left( C_{aaij} + C_{bbij} \right) 
	\left( C_{abij} -\frac{1}{2} \left( C_{aaij} + C_{bbij} \right) \right) \right)
\end{align}

The 2-body reduced density matrix has many different types of element. 
\begin{align}
	d_{aaaa} = 2 \sum_{i (\neq a)} C_{aaii} C_{aaii} 
	+ \sum_{i<j (\neq a)} C_{aaij} C_{aaij}
\end{align}

\begin{align}
	d_{aaab} &= d_{aaba} = d_{abaa} = d_{baaa} \\
	&= \sum_{i (\neq a,b)} \left( C_{aaii} C_{iiab} - \frac{1}{2} C_{aabi} C_{bbai} \right) 
	+ \frac{1}{2} \sum_{i<j (\neq a,b)} C_{aaij} \left( C_{abij} - \frac{1}{2} \left( C_{aibj} + C_{ajbi} \right) \right)
\end{align}

\begin{align}
	d_{aabb} &= d_{bbaa} \\
	&= 4 C_{aabb}C_{aabb} 
	+ \sum_{i (\neq a,b)} \left( C_{aabi} C_{aabi} + C_{bbai} C_{bbai} + \frac{1}{2} C_{iiab} C_{iiab} \right) \nonumber \\
	&+ \frac{1}{4} \sum_{i<j (\neq a,b)} \left( C_{abij}C_{abij} + C_{aibj}C_{aibj} + C_{ajbi}C_{ajbi} \right) \nonumber \\
	&- \frac{1}{4} \sum_{i<j (\neq a,b)} \left( C_{abij}C_{aibj} + C_{abij}C_{ajbi} + C_{aibj}C_{ajbi} \right)
\end{align}

\begin{align}
	d_{abba} &= d_{baab} \\
	 &= -2 C_{aabb} C_{aabb} 
	- \frac{1}{2} \sum_{i (\neq a,b)} \left( C_{aabi}C_{aabi} + C_{bbai}C_{bbai} - C_{iiab}C_{iiab} \right) \nonumber \\
	&+ \frac{1}{4} \sum_{i<j (\neq a,b)} \left( C_{abij} \left( C_{abij} - C_{aibj} - C_{ajbi} \right)
	- \frac{1}{2} C_{aibj}C_{aibj} - \frac{1}{2} C_{ajbi}C_{ajbi} + 2 C_{aibj}C_{ajbi} \right)
\end{align}

\begin{align}
	d_{abab} &= d_{baba} \\
	 &= 2 \sum_{i (\neq a,b)} C_{aaii} C_{bbii} + \sum_{i<j (\neq a,b)} C_{aaij} C_{bbij}
\end{align}

\begin{align}
	d_{aabc} &= d_{aacb} = d_{bcaa} = d_{cbaa} \\
	&= 2 C_{aabc} \left( C_{aabb} + C_{aacc} \right) - \frac{1}{2} C_{bbac} C_{ccab} 
	+ \sum_{i (\neq a,b,c)} \left( C_{aabi} C_{aaci} + \frac{1}{2} C_{iiab} C_{iiac} \right) \nonumber \\
	&+ \frac{1}{2} \sum_{i (\neq a,b,c)} \left( C_{bbai} + C_{ccai} \right) 
	\left( C_{aibc} - \frac{1}{2} C_{acbi} - \frac{1}{2} C_{abci} \right)
\end{align}

\begin{align}
	d_{abca} &= d_{baac} = d_{caab} = d_{acba} \\
	&= - C_{aabc} \left( C_{aabb} + C_{aacc} \right) + C_{bbac} C_{ccab} 
	+ \frac{1}{2} \sum_{i (\neq a,b,c)} \left( C_{iiab}C_{iiac} - C_{aabi}C_{aaci} \right) \nonumber \\
	&+ \frac{1}{2} \sum_{i (\neq a,b,c)} \left( C_{bbai} \left( C_{abci} - \frac{1}{2} C_{acbi} - \frac{1}{2} C_{aibc} \right)
	+ C_{ccai} \left( C_{acbi} - \frac{1}{2} C_{abci} - \frac{1}{2} C_{aibc} \right) \right)
\end{align}

\begin{align}
	d_{abac} &= d_{baca} = d_{acab} = d_{caba} \\
	&= - C_{aabc} C_{bbcc} + \sum_{i (\neq a,b,c)} \left( C_{aaii} C_{iibc} - \frac{1}{2} C_{bbci} C_{aabi} - \frac{1}{2} C_{ccbi}C_{aaci} \right)
\end{align}

\begin{align}
	d_{abcd} &= d_{cdab} = d_{dcba} = d_{bacd} \\
	&= C_{bbcd}C_{ccab} + C_{aacd}C_{ddbc} - \frac{1}{2} C_{bbad}C_{aabc} - \frac{1}{2} C_{ccad}C_{ddbc} \nonumber \\
	&+ \left( C_{aacc} + C_{bbdd} \right) \left( C_{abcd} - \frac{1}{2} C_{acbd} - \frac{1}{2} C_{adbc} \right)
\end{align}

\section{Paldus isomers for $a=1.2a_0$ and $a=1.6a_0$}
\label{sec:paldus}
\begin{figure}
	\begin{subfigure}{\textwidth}
		\includegraphics[width=0.485\textwidth]{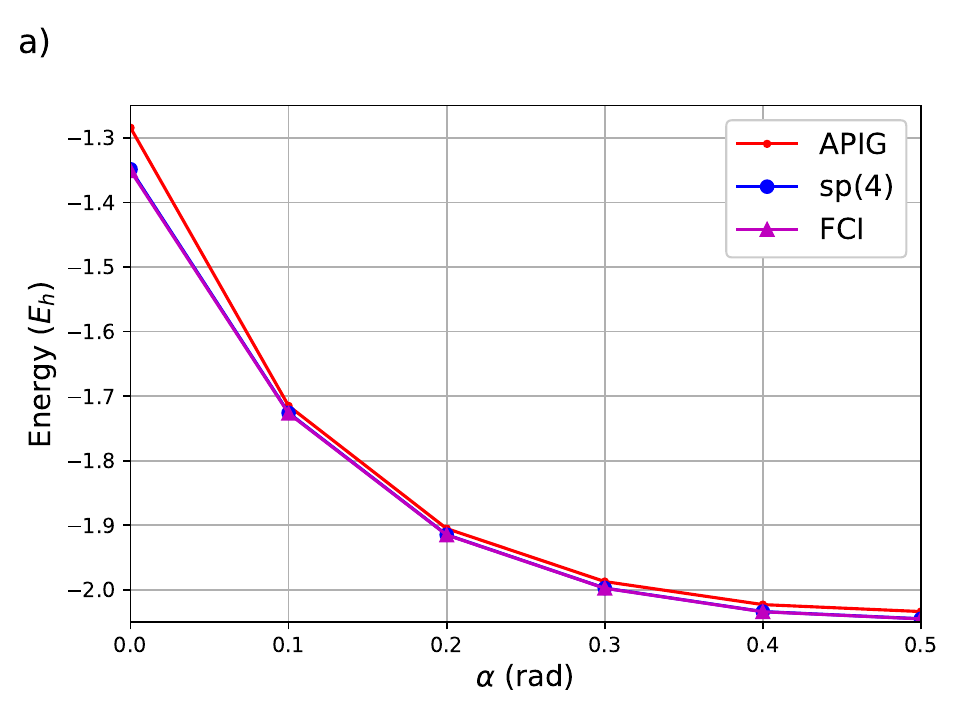}  \hfill
		\includegraphics[width=0.485\textwidth]{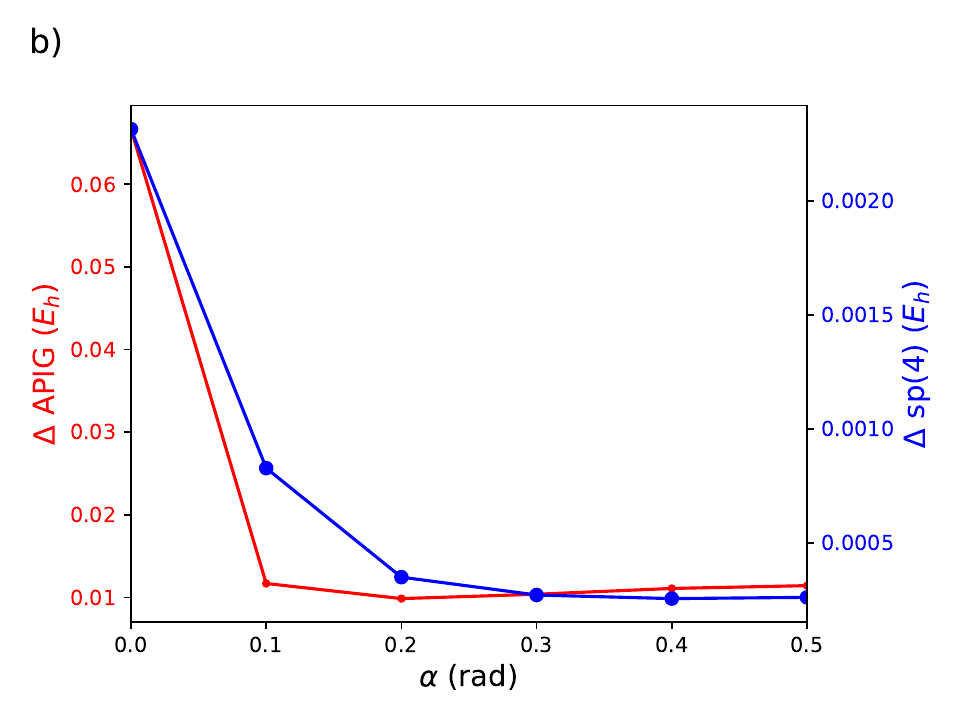}				
	\end{subfigure}
	\caption{Variational treatment of Paldus H4 (ring-opening of square to linear H$_4$) for $a=1.2a_0$ in the STO-6G basis: (a) sp(4), APIG, and FCI energies. (b) Error of sp(4) and APIG with respect to FCI. APIG results are computed in the basis of OO-DOCI orbitals.\cite{johnson:2023}}
	\label{fig:open_H4_curves_1p2}
\end{figure}

\begin{figure}
	\begin{subfigure}{\textwidth}
		\includegraphics[width=0.485\textwidth]{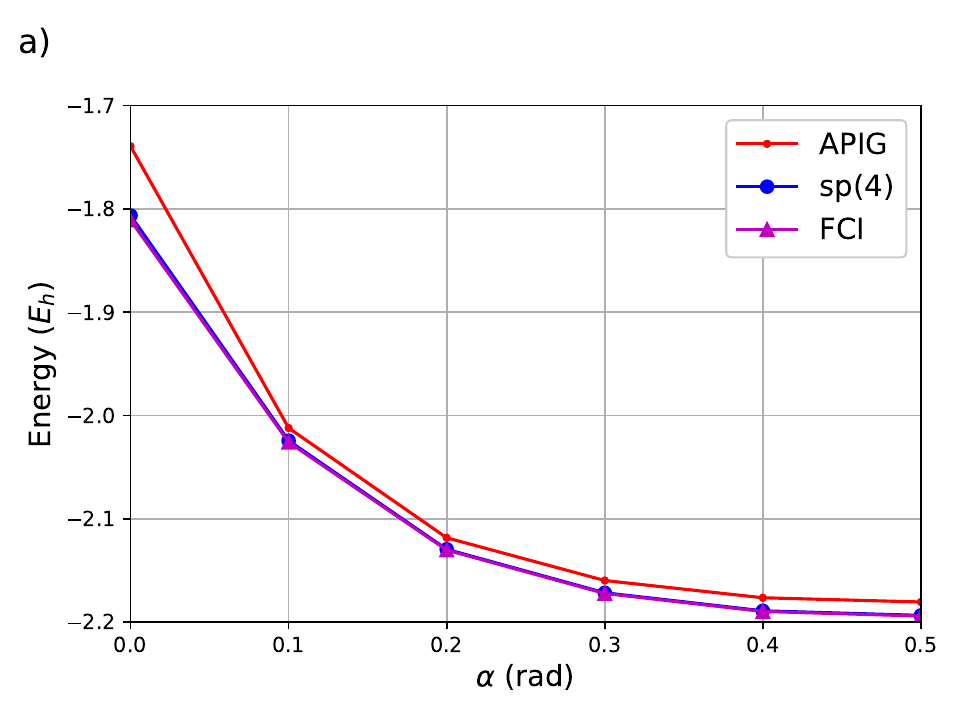}  \hfill
		\includegraphics[width=0.485\textwidth]{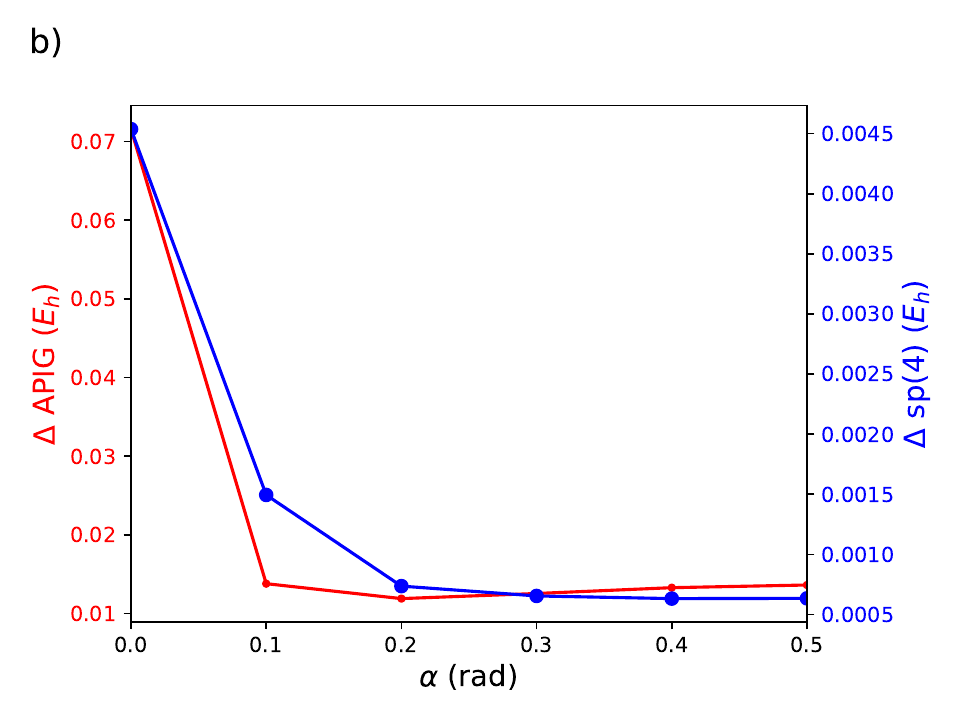}				
	\end{subfigure}
	\caption{Variational treatment of Paldus H4 (ring-opening of square to linear H$_4$) for $a=1.6a_0$ in the STO-6G basis: (a) sp(4), APIG, and FCI energies. (b) Error of sp(4) and APIG with respect to FCI. APIG results are computed in the basis of OO-DOCI orbitals.\cite{johnson:2023}}
	\label{fig:open_H4_curves_1p6}
\end{figure}

\begin{figure}
	\begin{subfigure}{\textwidth}
		\includegraphics[width=0.485\textwidth]{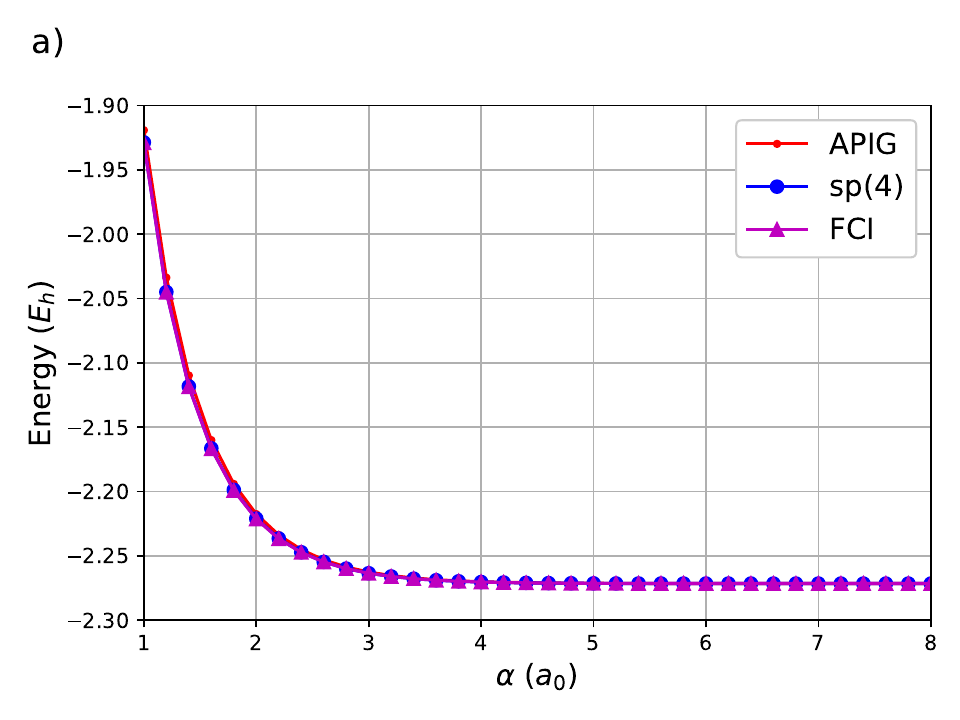}  \hfill
		\includegraphics[width=0.485\textwidth]{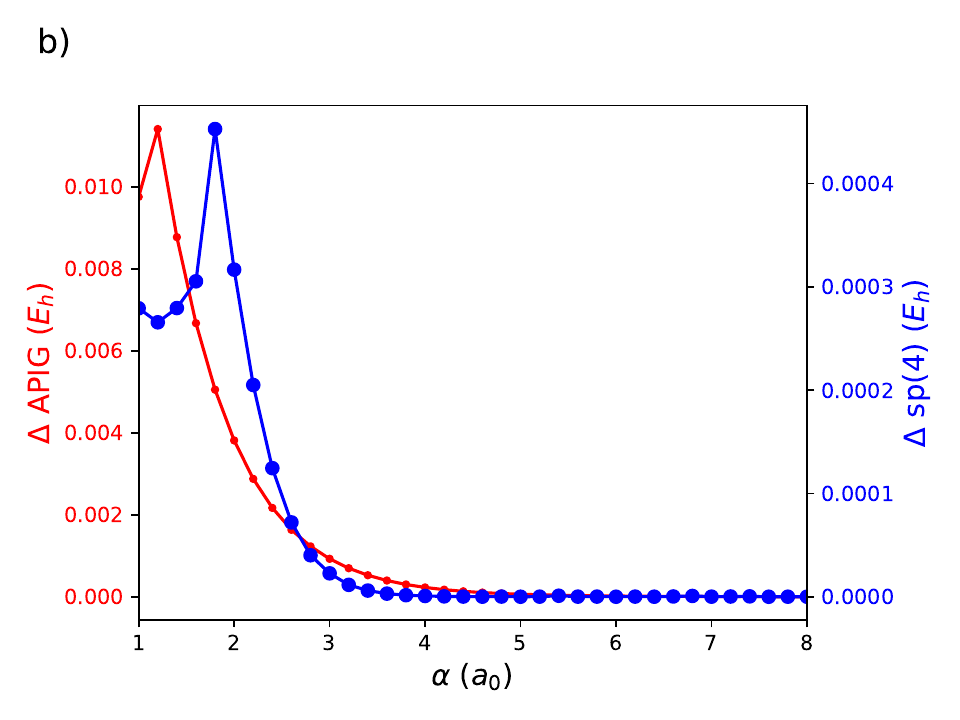}				
	\end{subfigure}
	\caption{Variational treatment of Paldus D4 for $a=1.2a_0$ in the STO-6G basis: (a) sp(4), APIG, and FCI energies. (b) Error of sp(4) and APIG with respect to FCI. APIG results are computed in the basis of OO-DOCI orbitals.\cite{johnson:2023}}
	\label{fig:D4_curves_1p2}
\end{figure}

\begin{figure}
	\begin{subfigure}{\textwidth}
		\includegraphics[width=0.485\textwidth]{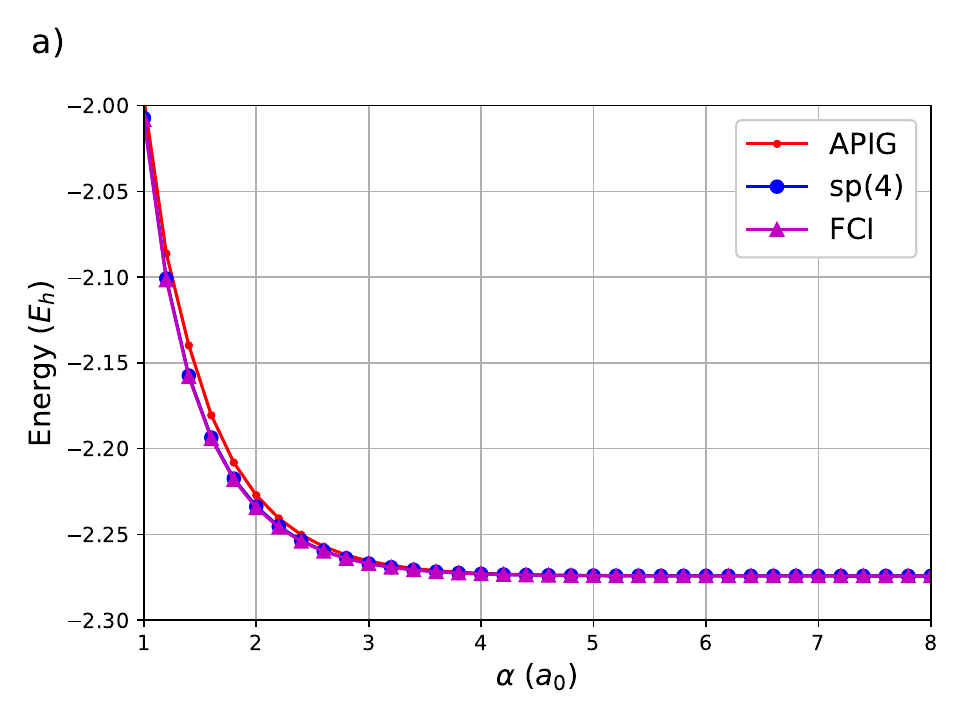}  \hfill
		\includegraphics[width=0.485\textwidth]{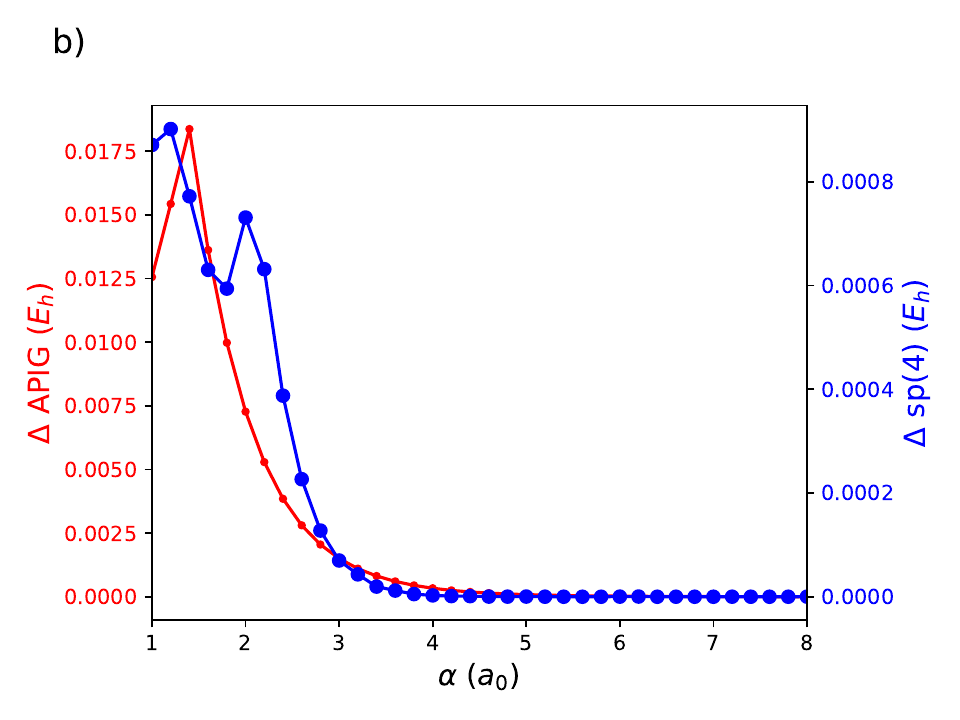}				
	\end{subfigure}
	\caption{Variational treatment of Paldus D4 for $a=1.6a_0$ in the STO-6G basis: (a) sp(4), APIG, and FCI energies. (b) Error of sp(4) and APIG with respect to FCI. APIG results are computed in the basis of OO-DOCI orbitals.\cite{johnson:2023}}
	\label{fig:D4_curves_1p6}
\end{figure}

\begin{figure}
	\begin{subfigure}{\textwidth}
		\includegraphics[width=0.485\textwidth]{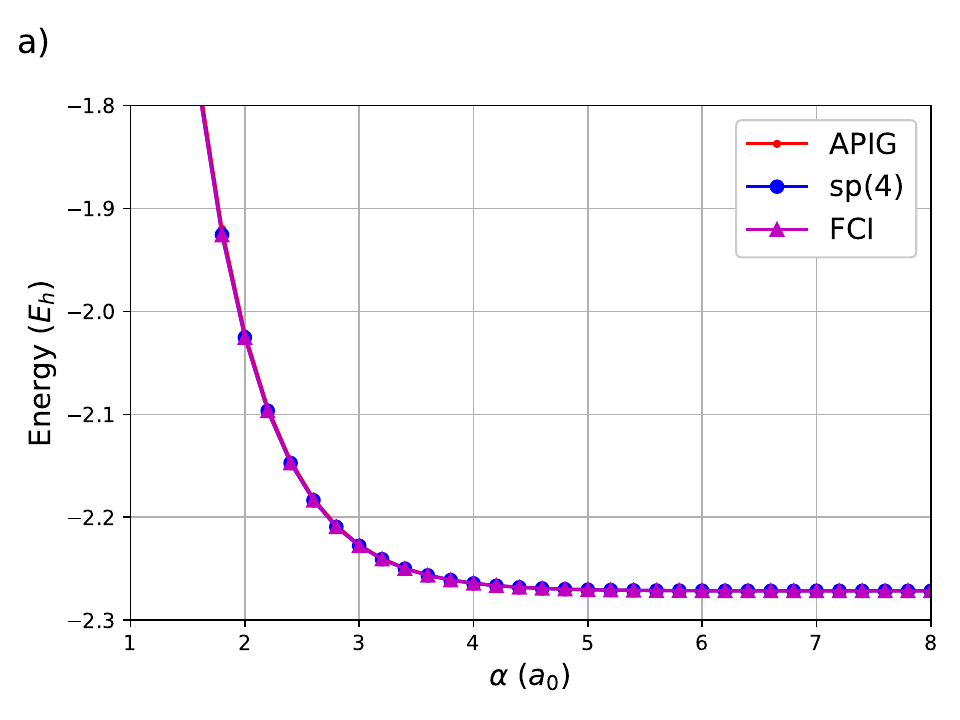}  \hfill
		\includegraphics[width=0.485\textwidth]{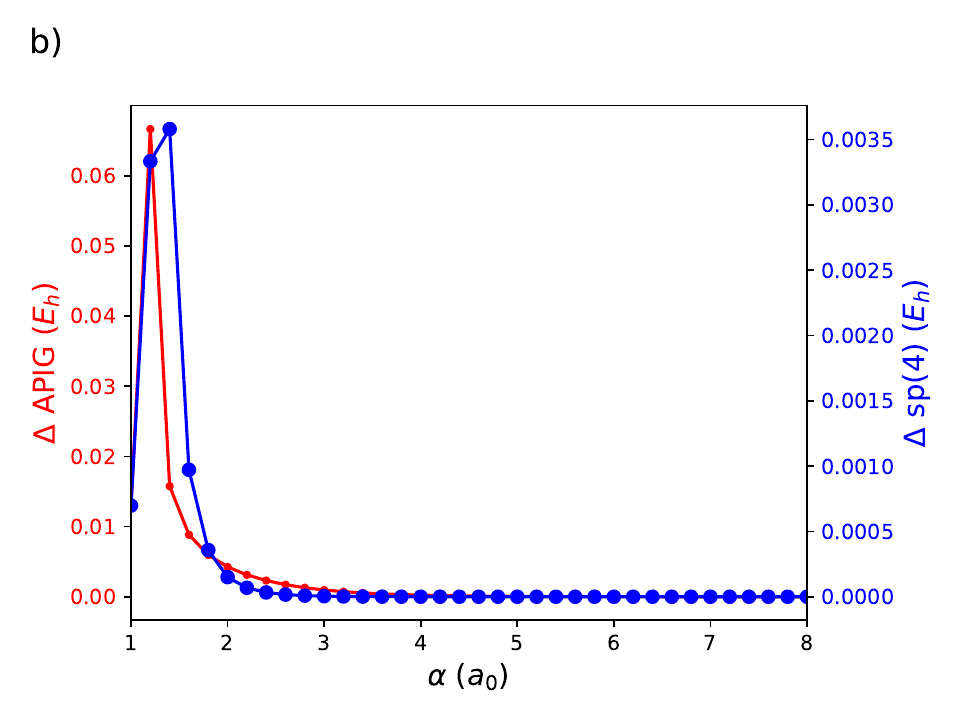}				
	\end{subfigure}
	\caption{Variational treatment of Paldus P4 for $a=1.2a_0$ in the STO-6G basis: (a) sp(4), APIG, and FCI energies. (b) Error of sp(4) and APIG with respect to FCI. APIG results are computed in the basis of OO-DOCI orbitals.\cite{johnson:2023}}
	\label{fig:P4_curves_1p2}
\end{figure}

\begin{figure}
	\begin{subfigure}{\textwidth}
		\includegraphics[width=0.485\textwidth]{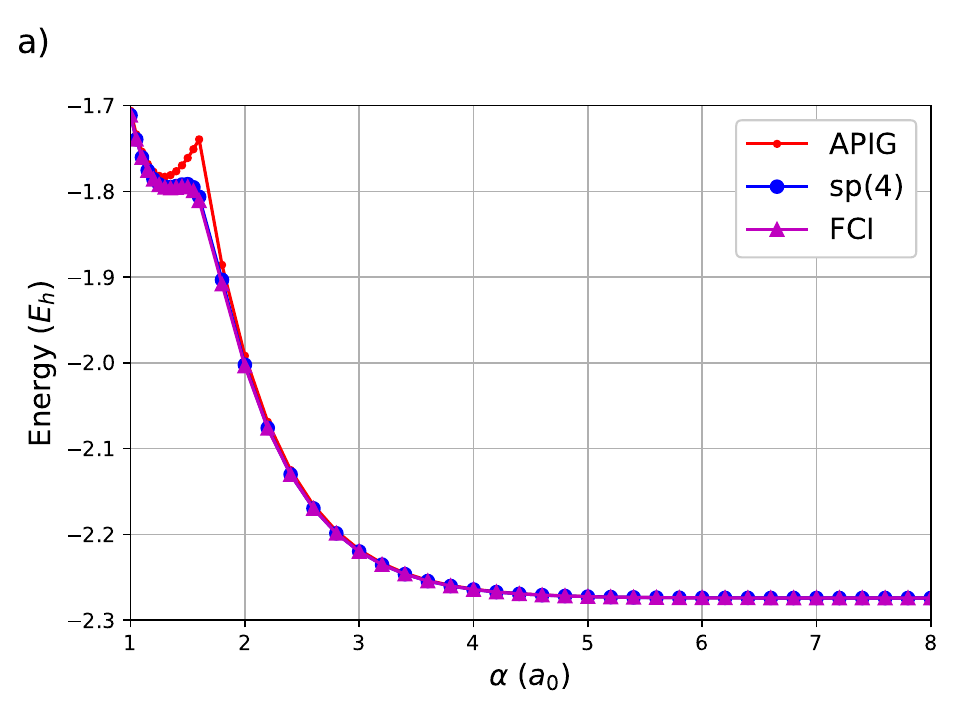}  \hfill
		\includegraphics[width=0.485\textwidth]{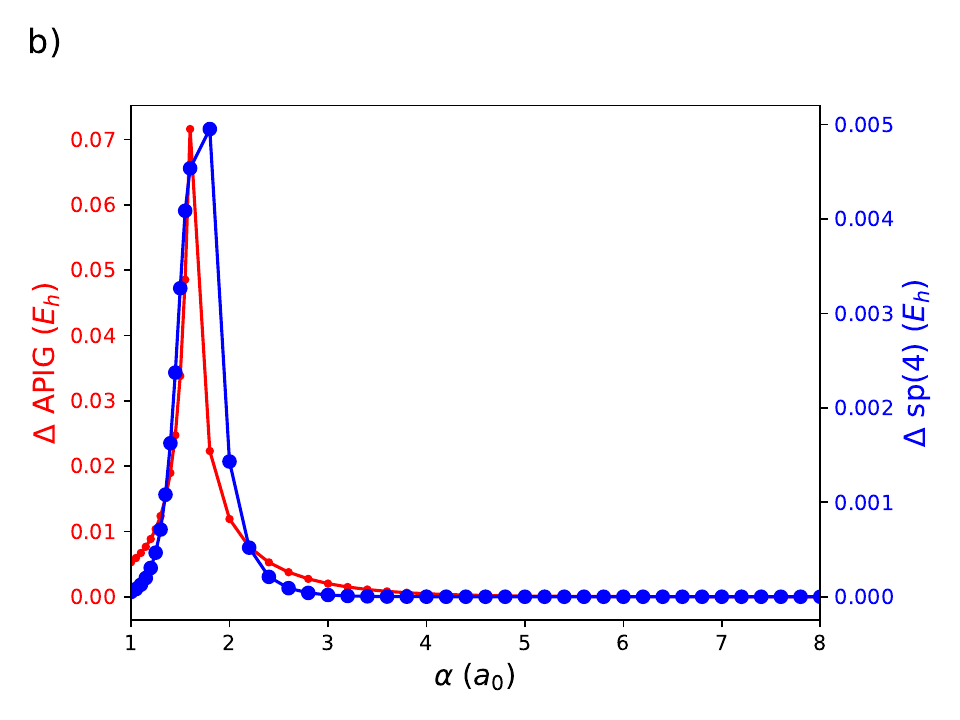}				
	\end{subfigure}
	\caption{Variational treatment of Paldus P4 for $a=1.6a_0$ in the STO-6G basis: (a) sp(4), APIG, and FCI energies. (b) Error of sp(4) and APIG with respect to FCI. APIG results are computed in the basis of OO-DOCI orbitals.\cite{johnson:2023}}
	\label{fig:P4_curves_1p6}
\end{figure}

\bibliography{lib_spn}

\bibliographystyle{unsrt}

\end{document}